\documentclass[12pt,preprint]{aastex}

\shorttitle{2008 April 9 Cartwheel Flare}
\shortauthors{Savage, McKenzie, Reeves, Forbes, and Longcope}

\begin{document}


\title{Reconnection Outflows and Current Sheet Observed with Hinode/XRT in the 2008 April 9 ``Cartwheel CME" Flare}


\author{$^{1}$Sabrina L. Savage, $^{1}$David E. McKenzie, $^{2}$Katharine K. Reeves, \\ $^{3}$Terry G. Forbes, $^{1}$Dana W. Longcope}
\affil{$^{1}$Department of Physics, Montana State University, PO Box 173840, Bozeman, MT 59717-3840, USA}
\affil{$^{2}$Harvard-Smithsonian Center for Astrophysics, 60 Garden Street MS 58, Cambridge, MA  02138}
\affil{$^{3}$Institute for the Study of Earth, Oceans, and Space (EOS), University of New Hampshire, 39 College Road, Durham, NH  03824}


\begin{abstract}

Supra-arcade downflows (SADs) have been observed with \textit{Yohkoh}/SXT (soft X-rays (SXR)), \textit{TRACE} (extreme ultra-violet (EUV)), \textit{SoHO}/LASCO (white light), \textit{SoHO}/SUMER (EUV spectra), and \textit{Hinode}/XRT (SXR).  Characteristics such as low emissivity and trajectories which slow as they reach the top of the arcade are consistent with post-reconnection magnetic flux tubes retracting from a reconnection site high in the corona until they reach a lower-energy magnetic configuration.  Viewed from a perpendicular angle, SADs should appear as shrinking loops rather than downflowing voids.  We present XRT observations of supra-arcade downflowing loops (SADLs) following a coronal mass ejection (CME) on 2008 April 9 and show that their speeds and decelerations are consistent with those determined for SADs.  We also present evidence for a possible current sheet observed during this flare that extends between the flare arcade and the CME.  Additionally, we show a correlation between reconnection outflows observed with XRT and outgoing flows observed with LASCO.

\end{abstract}


\keywords{Magnetic reconnection --- Sun: corona --- Sun: flares --- Sun: coronal mass ejections (CMEs) --- Sun: magnetic topology --- Sun: X-rays}

\section{Introduction}

While the details of flare dynamics are debatable, the general energy release mechanism is widely accepted to arise from magnetic reconnection.  Direct evidence of reconnection, however, has been scarce and often questionable.  In all models of reconnection, loops flowing both towards and away from the reconnection site are an inevitable theoretical consequence (\citeauthor{carm} \citeyear{carm}; \citeauthor{sturrock} \citeyear{sturrock}; \citeauthor{hirayama} \citeyear{hirayama}; \citeauthor{kopp} \citeyear{kopp}).  Observationally, though, these loops require very special circumstances in order to image since they are likely to be relatively devoid of emitting plasma and form high in the corona where they are viewed against a dark background.  Supra-arcade downflows (SADs) have been observed in several flares and interpreted as the cross-sections of these shrinking loops as they retract through a bright fan (\citeauthor{mck99} \citeyear{mck99}; \citeauthor{mck00} \citeyear{mck00}; \citeauthor{sadsI} \citeyear{sadsI}).  Cusped flares, like those predicted by the standard models, have been shown to have signatures of retracting loops (\citeauthor{forbesacton96} \citeyear{forbesacton96}; \citeauthor{reeves08} \citeyear{reeves08}).  Imaging individual loops retracting above the flaring site with high enough temporal and spatial resolution has proven to be a challenge due to observational limitations.  In order to observe the downflows which occur above the post-eruption arcade, the flare must occur near the limb; and long image exposures, which inevitably saturate the flaring site and are therefore not desirable for most flare observations, must be taken to provide proper contrast in the low signal to noise region above the arcade.

Coronal mass ejections (CMEs) are frequently observed to be associated with eruptive flares.  Current sheets are expected to extend between the arcade region and the CME \citep{forbesacton96}.  While the current sheets themselves are probably too narrow to be fully resolved with current instrumentation, calculations have shown that conduction fronts lead to the formation of a sheath of hot plasma surrounding the current sheet that widens the observable structure (\citeauthor{reevesinpress} \citeyear{reevesinpress}; \citeauthor{seaton09} \citeyear{seaton09}; \citeauthor{yokshib98} \citeyear{yokshib98}).  Recent modeling has shown that this hot plasma sheath can be observed by a sensitive X-ray imager such as Hinode's X-ray Telescope (XRT) (\citeauthor{reevesinpress} \citeyear{reevesinpress}).  Current sheet observations have been claimed and analyzed for several flares using EUV and white light coronagraphs (\citeauthor{ciaray08} \citeyear{ciaray08}; \citeauthor{linetal07} \citeyear{linetal07}; \citeauthor{webb03} \citeyear{webb03}; \citeauthor{ciaetal02} \citeyear{ciaetal02}; \citeauthor{ko02} \citeyear{ko02}).  Because white-light coronagraphs measure polarization brightness, which is directly related to density, these measurements indicate that the density in the structures surrounding the current sheet is elevated compared with the background corona.

In the following sections, we describe the ``Cartwheel CME" flare as seen by XRT and LASCO.  We describe the XRT observations in detail, which include a candidate current sheet, shrinking loops, and flows, and then show correspondences between XRT and LASCO flows.  We also discuss a possible scenario for interpreting the observations based on magnetic modeling. 

\section{Observations}

\begin{figure}[!ht] 
\begin{center}
\includegraphics[width=1\textwidth]{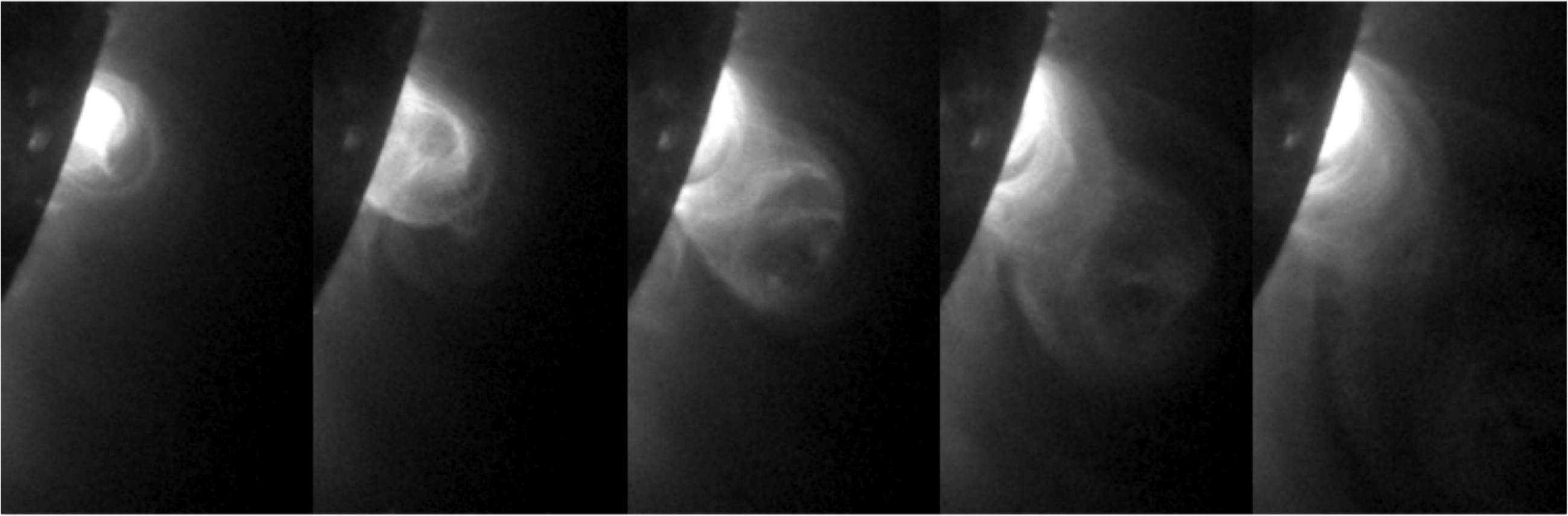}
\caption{A CME is observed by XRT between 09:16 and 10:11 UT.}
\label{CME}
\end{center}
\end{figure}

The stringent conditions required to observe faint structures associated with reconnection above a flaring region are all met during a flare observed by Hinode's XRT on 2008 April 9 between 09:16 UT and 17:32 UT in association with active region 10989 (see \citeauthor{golub07} \citeyear{golub07} for a description of the XRT instrument).  This flare has colloquially been nicknamed the ``Cartwheel CME" flare because the observed structure, which we interpret as a flux rope in Section 4, appears to rotate as it is ejected from the Sun.   The region is approximately 23$^{\circ}$ behind the west limb.  Several instruments observed this event including XRT, \textit{STEREO A}/SECCHI, TRACE, and LASCO.  The flare is observed on the limb within SECCHI's field of view (FOV), and from these observations, we place the actual flare start time at approximately 08:53 UT.  This paper will focus on observations from XRT and LASCO with support from SECCHI.  XRT's images are taken using the Al/poly filter with 1 - 16 second exposures.  There is no GOES signature for this flare due to its occurrence behind the limb; however, this fortunate observational situation offers a unique, relatively deep-exposure look at the supra-arcade region as the limb occulted the bright footpoints.  

\begin{figure}[!ht] 
\begin{center}
\includegraphics[width=0.3\textwidth]{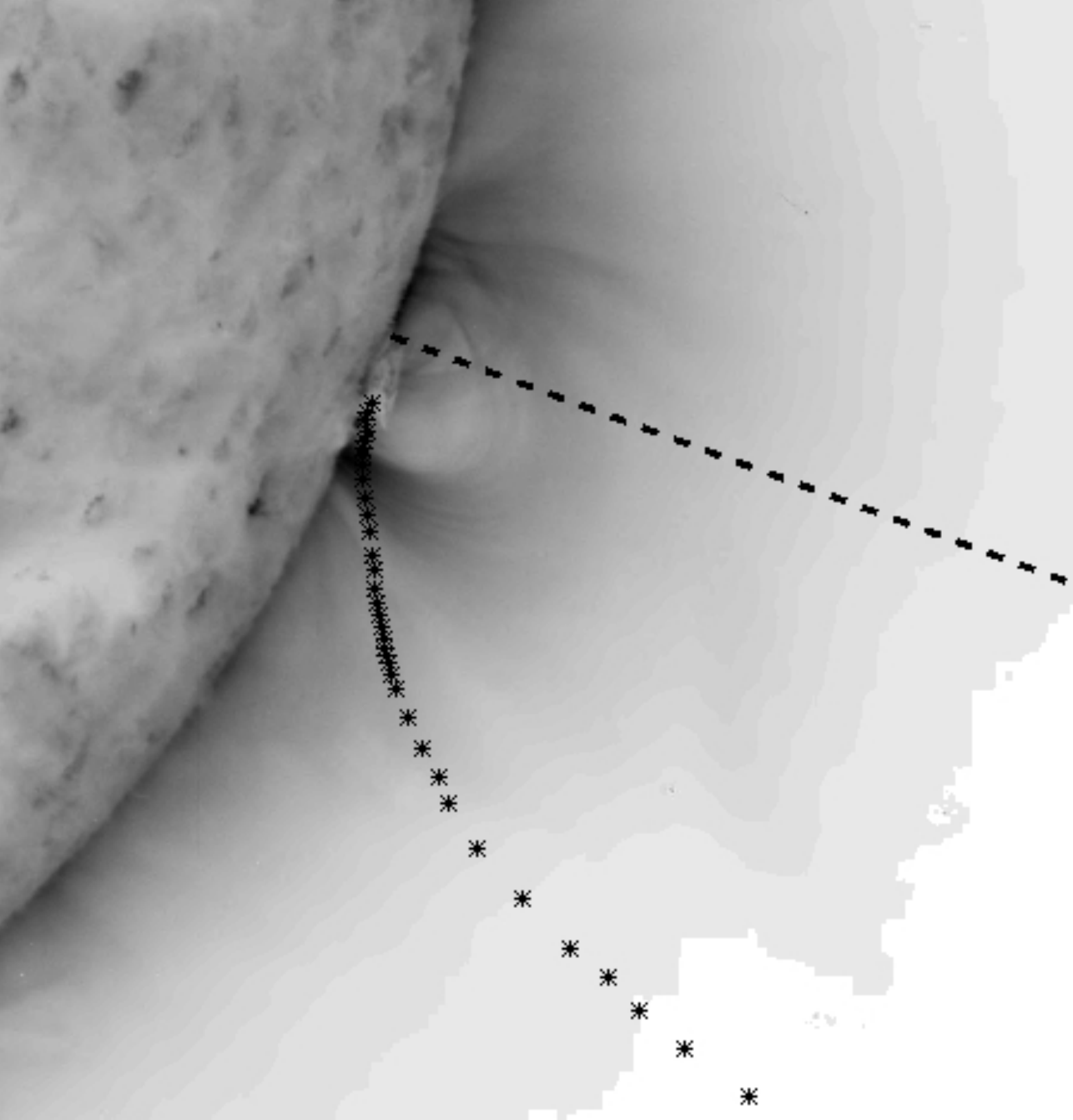}
\includegraphics[width=0.66\textwidth]{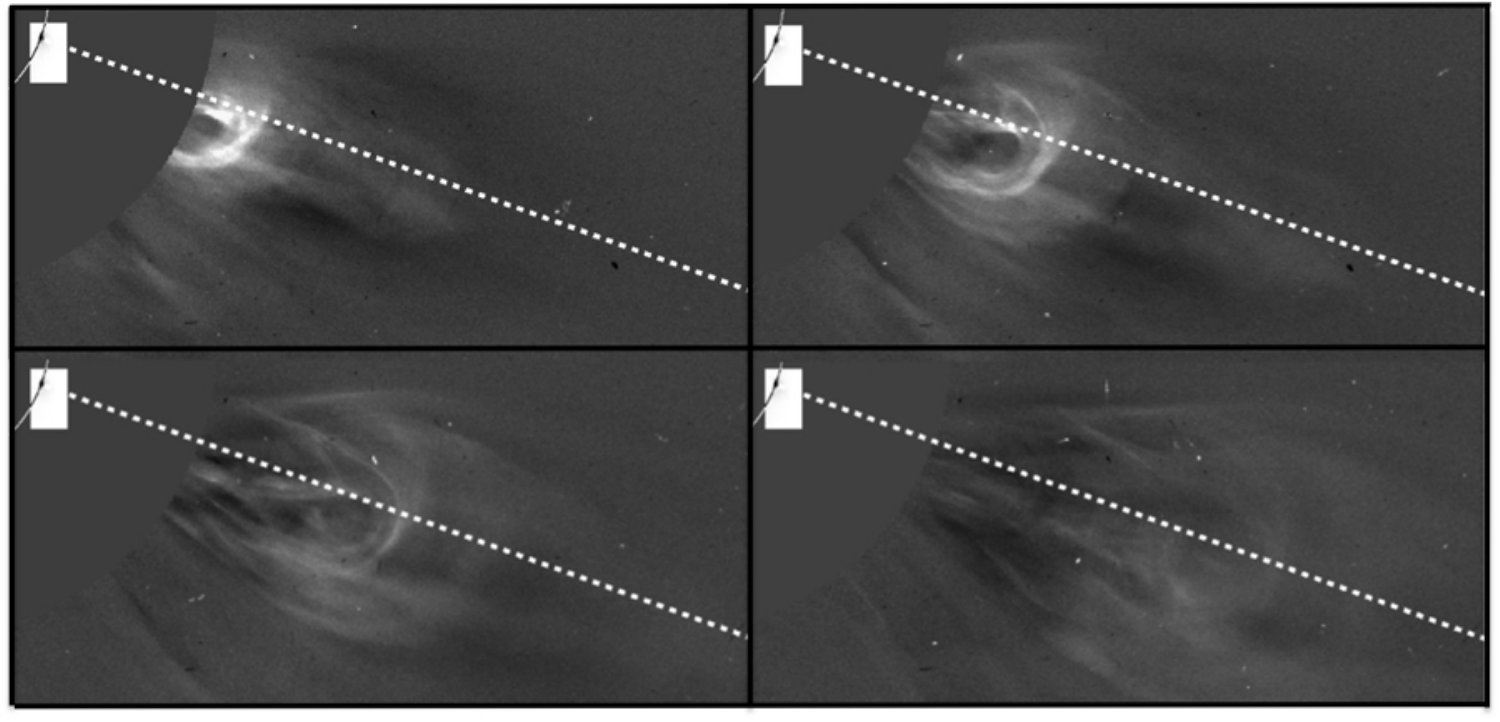}
\caption{\textit{Left}:  Path of the erupted filament as seen by \textit{STEREO A}/SECCHI.  The dashed line indicates the radial direction extending from the active region as seen on the limb.  The curved path follows the trajectory of the erupted structure from 08:53 UT (time of the reverse-scaled background image).  \textit{Right}:  LASCO C2 frames (11:06 UT, 11:26 UT, 11:50 UT, 12:26 UT) showing the CME path beyond $\sim$2.5 R$_{\odot}$.  The dashed line indicates the radial direction extending from the active region projected onto the plane of the sky.  The XRT FOV is shown in the upper left-hand corner of each image.}
\label{cme_path}
\end{center}
\end{figure}

A large body of EUV- and X-ray-emitting mass is observed by XRT from 09:16 UT to 10:11 UT (Figure~\ref{CME}).  The flare is obscured by the limb up to about 70 Mm in the XRT FOV.  The speed of the structure as observed by XRT increases from $\sim$80 to $\sim$180 km s$^{-1}$.  A white-light CME enters the SoHO/LASCO FOV at 11:06 UT and proceeds through the LASCO C2 FOV with an average speed of $\sim$450 km s$^{-1}$ \citep{landi10}.  The onset of the filament eruption is observed by SECCHI beginning at about 08:53 UT.  Figure~\ref{cme_path} (left) depicts the curved path of the eruption within the SECCHI FOV.  Figure~\ref{cme_path} (right) shows the CME as it passes through the LASCO C2 FOV.  The dashed white line indicates the radial direction extending from the active region projected onto the plane of the sky.  These observations indicate that the erupted structure initially moves in a non-radial direction toward the observer with its path becoming more radial as it approaches $\sim$2.5 R$_{\odot}$.  The deflection may be the result of interaction with an open field region to the southeast of the active region, though such interaction remains speculative (Section 4).  (See \cite{landi10} for a detailed analysis of this CME.)  The observations described in the following sections all occurred after the CME, and in the region from which it departed.

\section{Analysis}

The original XRT images clearly show loops shrinking towards a bright arcade region.  Sharpening the images improves the visibility of the individual loops at greater heights and makes it possible to detect faint, moving features previously unnoticed in the unsharpened movies.

The 358 original images examined between 08:01 and 17:32 UT are taken using an automatic exposure control setting.  This resulted in some dynamic exposure times with a typical cadence of one minute.  The median exposure setting is nearly 6 seconds which is about 60 times longer than exposure durations taken with the Al/poly filter of similar unocculted active regions on the verge of saturation (e.g. AR 10978; December 18, 2007). There is a large data gap of 155 minutes occurring between 13:47 and 16:22 UT.  Despite normalizing, the widely varying exposure times make comparing successive images difficult due to the varying low signal-to-noise background level; therefore, the first step in sharpening is to eliminate the occurrence of non-repeating exposure durations within a time span of 30 minutes.  The exposure-filtered set is flattened by dividing by the mean of an array of temporally-adjacent images with similar exposure durations.  Then to enhance movement, each image is running-mean-differenced by subtracting the mean of the same set of similar surrounding images.  Finally, the image set is byte-scaled in order to reduce any residual flickering (i.e. all of the pixels per image are binned based on a 0 to 255 intensity scale which reduces the dynamic scaling between images of different exposure lengths).  The resulting image set has 274 images with typical cadences ranging from 1 to 5 minutes between data gaps (median:  1 minute).  

\subsection{Candidate Current Sheet} 

\begin{figure}[!ht] 
\begin{center}
\includegraphics[width=0.65\textwidth]{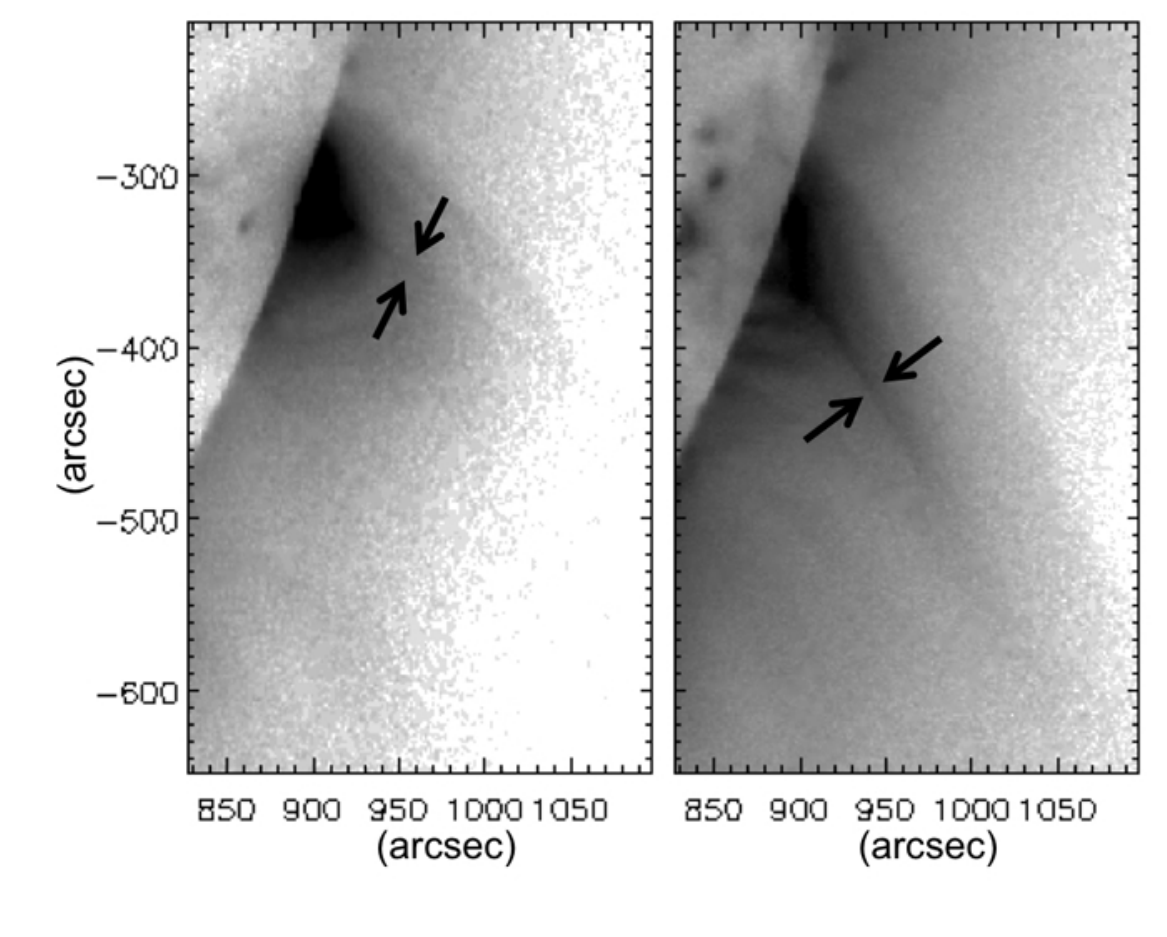}
\caption{\textit{Left}: (11:08:24 UT)  \textit{Right}: (17:31:55 UT)  These reverse-scale images highlight the bright, thin linear feature (i.e. the candidate current sheet) extending from the apex of the arcade region from the 2008 April 9 event.  Note the southward drift of approximately 4 deg/hr.  All flows track along this feature even as it progresses southward.}
\label{cs_rotation}
\end{center}
\end{figure}

The sharpened image set reveals flows moving both towards and away from the solar surface (henceforth referred to in the text as downflows and upflows respectively) and shrinking loops that can be tracked after the CME has left the XRT FOV.  All of these flows both toward and away from the Sun, including the apexes of the shrinking loops, follow the direction of a bright, thin linear feature which extends from the apex of the arcade region.  This feature becomes apparent at 11:00 UT following a nearly 40 minute data gap in the image sequence and slowly progresses southward about 25$^{\circ}$ over 6.5 hours for an average rate of approximately 4 deg/hr (see Figure~\ref{cs_rotation}).  The rotation rate at a latitude of $-$18$^{\circ}$, the approximate location of the flare, is about 0.6 deg/hr; therefore, the solar differential rotation cannot account for a projected drift of the linear feature. 

The appearance of this feature is similar to those interpreted as current sheets by \cite{ciaray08}, \cite{webb03}, and \cite{ko02}.  In particular, this feature is thin and bright compared to its surroundings and is located above the flare arcade.  In all cases, coronal magnetic fields cannot be measured, so the inference of a current sheet is circumstantial; however, the position of this feature at the top of the arcade and its orientation between the arcade and CME are consistent with the interpretation of a current sheet.  While the feature appears thin, the actual current sheet may be even thinner based on modeling which shows that current sheet structures can have surrounding areas of hot temperature due to conduction fronts, making the observable structure wider than the actual current sheet (\citeauthor{reevesinpress} \citeyear{reevesinpress}; \citeauthor{seaton09} \citeyear{seaton09}; \citeauthor{yokshib98} \citeyear{yokshib98}).  Recent modeling has shown that because of its sensitivity to high temperature plasma, XRT is able to observe this hot structure \citep{reevesinpress}.  An additional piece of circumstantial evidence in favor of the current sheet interpretation is the motion of the observed shrinking loops.  The shrinking loops observed in the XRT data display cusped looptops, and the apex of each loop tracks along this feature, consistent with the Kopp \& Pneuman model \citep{kopp}.

\begin{figure}[!ht] 
\begin{center}
\includegraphics[width=0.8\textwidth]{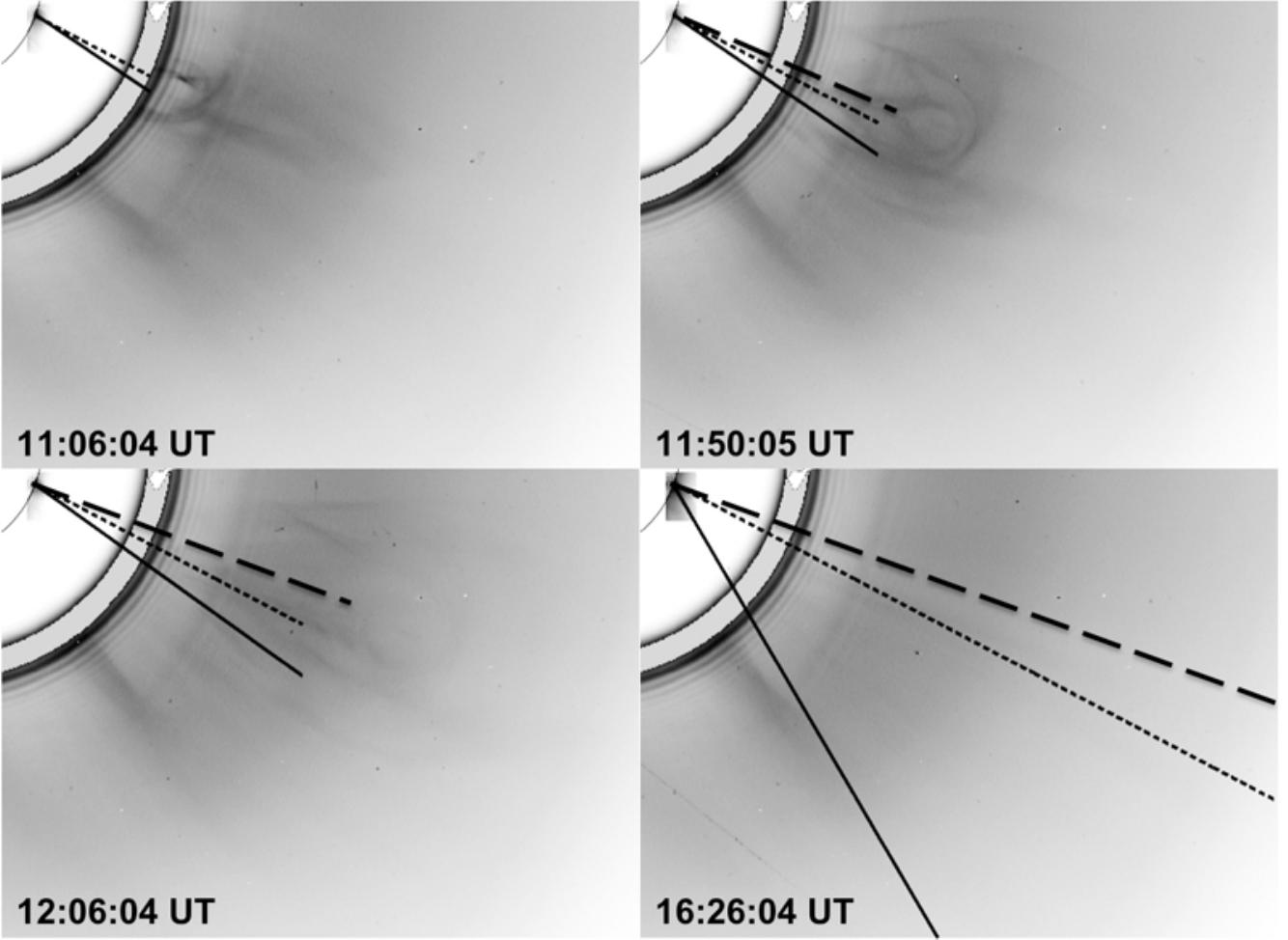}
\caption{Reverse-scale composite images with an XRT inset overlaid onto corresponding LASCO C2 images.  The candidate current sheet direction is represented by the solid line extending from the XRT arcade.  The length of the line is extended to near the base of the CME in the LASCO images.  The dotted line represents the initial direction of the CME as measured from the initial detection in the LASCO images (top left panel).  The CME path direction as manually identified in each subsequent LASCO image is indicated by the dashed line.  Notice that the CME does not follow a straight plane-of-sky path.  The times indicated are those of the LASCO images.}
\label{cs_rotation_composites}
\end{center}
\end{figure}

This feature (henceforth referred to as the candidate current sheet or CCS) is initially detected at a position angle $\sim$7 degrees southward of the CME when it first appears in the LASCO FOV at $\sim$2.5 R$_{\odot}$ around 11 UT.  It then immediately begins slowly drifting away from the LASCO CME path during its aforementioned apparent southward progression.  The alignment of the CCS with respect to the CME, and also the southward progression of the CCS, are consistent with the measurements from \cite{webb03} for 59 apparent current sheets.  Figure~\ref{cs_rotation_composites} displays reverse-scale composite images with an XRT inset overlaid onto corresponding LASCO C2 images.  The direction of the CCS is first determined in the full-scale XRT images and then expanded into the LASCO C2 FOV to show the relationship between the CME path and the CCS direction.  The length of the displayed CCS roughly corresponds to the distance between the top of the arcade in the XRT images and the central base of the CME in the LASCO set.  The initial CME direction (dotted lines) is displayed in all panels while the current CME location (dashed lines) is updated.  The growing angle between the two directions shows that the CME does not follow a straight plane-of-sky path.  This conclusion is also supported by the CME path as seen within the XRT FOV as well as by SECCHI images near the base of the flare (see Figure~\ref{cme_path}).  The deflection occurs prior to 3 R$_{\odot}$.

\begin{figure}[!ht] 
\begin{center}
\includegraphics[width=0.7\textwidth]{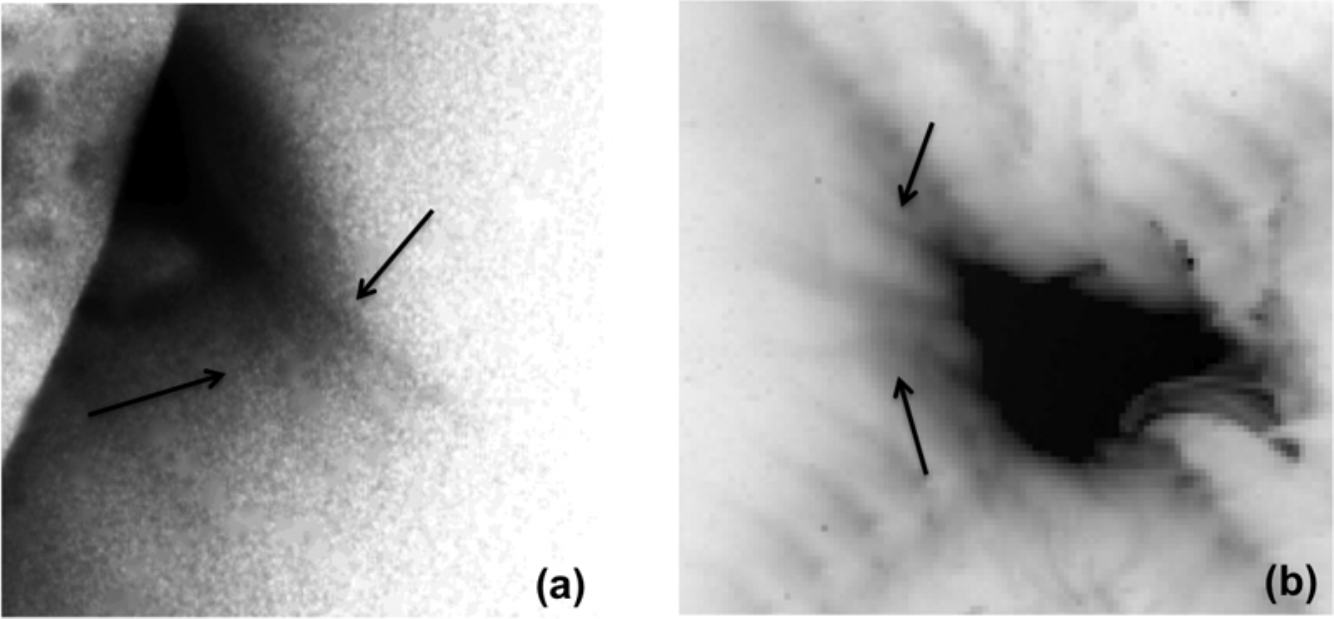}
\caption{(a)  (13:09 UT)  Reverse-scale XRT image highlighting the fan-like appearance of the candidate current sheet near the midpoint of the XRT image sequence.  This suggests that the CCS is not being viewed completely edge-on.  (b)  Reverse-scale TRACE image from July 23, 2002 at 00:40 UT.  The orientation of this flare is nearly mirrored to that of the ``Cartwheel CME" flare but with a less-severe projection.   A fan above the post-eruption arcade loops similar to that seen in (a) is easily discernible.}
\label{cs_fan}
\end{center}
\end{figure}

Examination of the source active region for this CME and flare indicates that the axis of the arcade is likely not oriented directly along the XRT line of sight due to the inclination of the polarity inversion line (see Section 4 for more details).  Thus the proposed current sheet is not observed directly edge-on, but with some projection (as in \citeauthor{ko02} \citeyear{ko02} and \citeauthor{ciaray08} \citeyear{ciaray08}).  Such a projection would be consistent with the observation that near the midpoint of the XRT sequence, the CCS does not appear to be a single ray, but exhibits more of a fanlike structure (Figure~\ref{cs_fan} (a)).  This structure is similar to those observed in the current sheet analyzed by \cite{ko02} (see Figures 5 \& 6 therein).  Numerous other authors have noted such fanlike or ``spikey" structures above arcades (e.g., \citeauthor{svestka98} \citeyear{svestka98} and \citeauthor{mck99} \citeyear{mck99}).  An example of this structure as seen in a flare occurring near the East limb on July 23, 2002 is given in Figure~\ref{cs_fan} (b).  The orientation of this flare is mirror opposite to that of the ``Cartwheel CME" flare but with a less-severe projection angle.

\begin{figure}[!ht] 
\begin{center}
\includegraphics[width=0.26\textwidth]{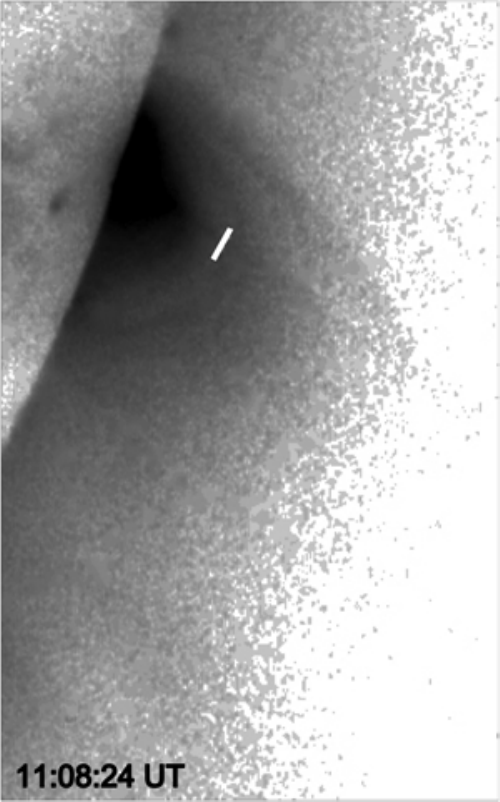}
\includegraphics[width=0.6\textwidth]{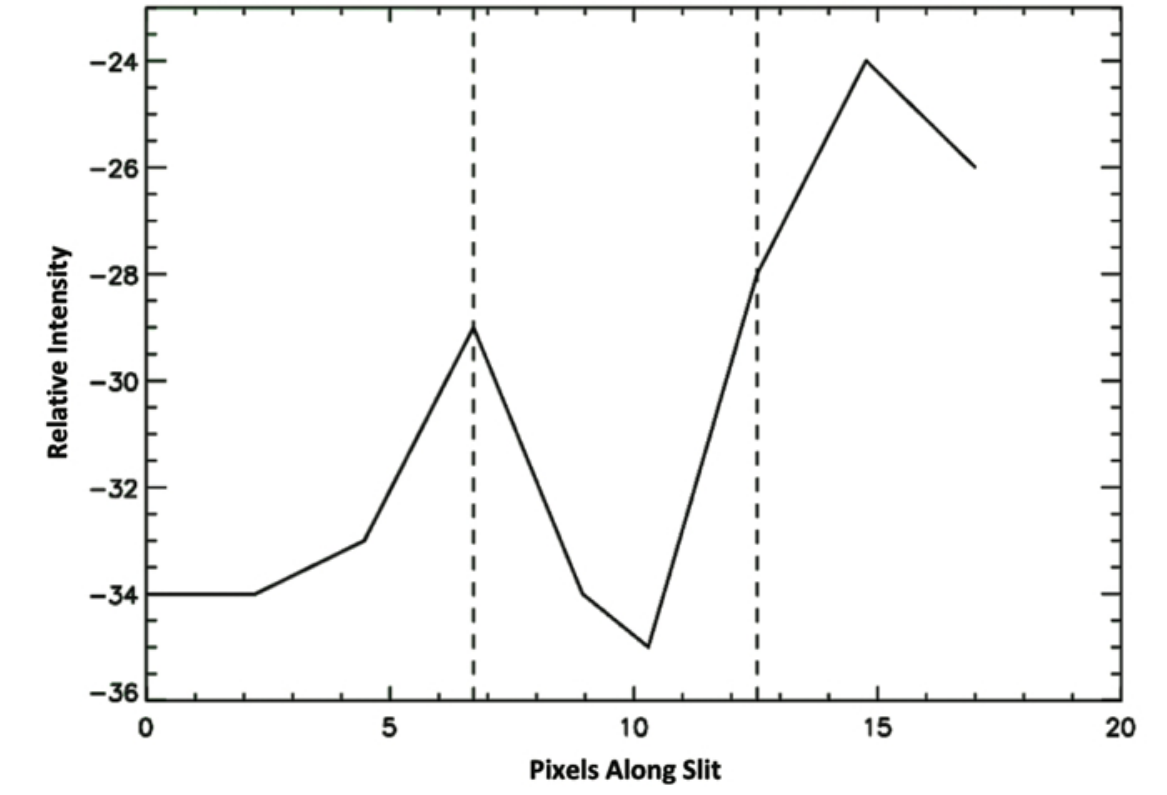}
\caption{\textit{Left:}  Example slice (white line) across the CCS for determining its thickness.  \textit{Right:}  The intensity profile across the slit.  The vertical dashed lines correspond to the positions chosen as the edges of the CCS.  Note that the scales of the image and the profile have been reversed.}
\label{cs_slice}
\end{center}
\end{figure}

\begin{figure}[!ht] 
\begin{center}
\includegraphics[width=0.25\textwidth]{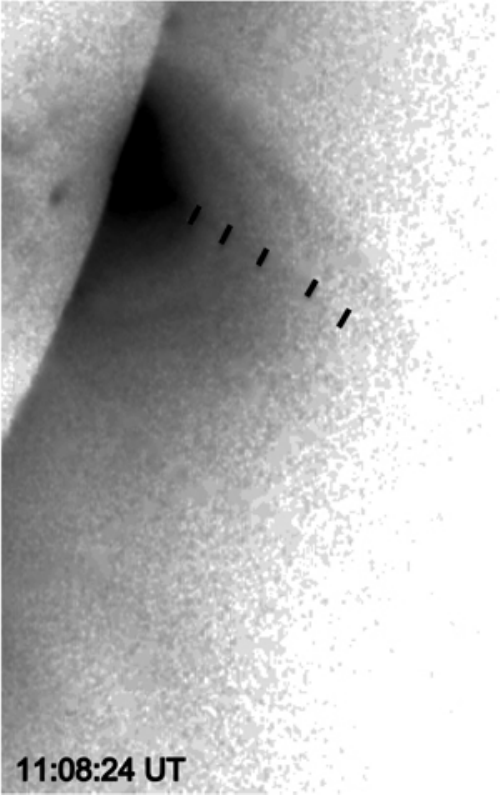}
\includegraphics[width=0.25\textwidth]{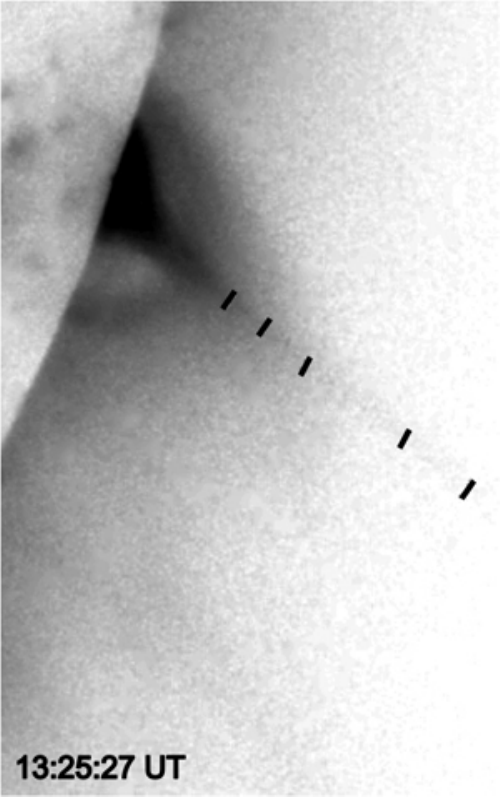}
\includegraphics[width=0.25\textwidth]{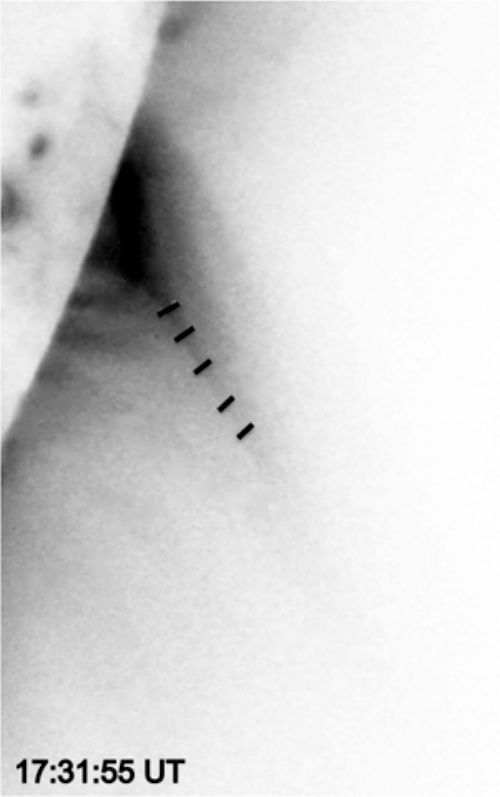}
\caption{Reverse-scaled images selected for determining the CCS thickness.  The slices across the CCS indicate the thickness determined for that position.}
\label{cs_thickness}
\end{center}
\end{figure}

The average thickness of the CCS (or rather the hot plasma surrounding the actual current sheet) is determined to be on order of (4-5)x10$^{3}$ km, depending on the image time.  This thickness is measured by extracting slits across the current sheet feature and using the slit intensity profile as a guide for determining the CCS thickness (see Figure~\ref{cs_slice} for an example slice and profile).  The selected images and the slits used to determine average thicknesses are displayed in Figure~\ref{cs_thickness}.  These thickness values should be regarded as upper limit values as they represent a thickness that is not being viewed perfectly edge-on (see Figure~\ref{cs_fan}).

\clearpage

\cite{ciaray08} obtained a CS thickness range of 30-60x10$^{3}$ km for the 2003 November 4 flare using UVCS and geometrical arguments.  They claim that the broadened line profiles obtained from UVCS measurements ``must result from either bulk flows or turbulence."  Furthermore, they predict the thickness due to turbulence to be $\leq$ 4x10$^{3}$ km.  The similarity between this prediction and the measurements obtained with XRT for the ``Cartwheel CME" flare suggest that turbulence may play a role in broadening the CS thickness.

Thickness estimates have been made by various authors for several other current sheets with similar orientations.  Our thickness estimate of (4-5)x10$^{3}$ km is at least 10 times narrower than the estimate range of (30-100)x10$^{3}$ km measured by \cite{ciaray08}, \cite{linetal07}, \cite{webb03}, \cite{ciaetal02}, and \cite{ko02}.  The discrepancy in thickness values may arise from the fact that we observe very hot plasma with XRT's Al/poly filter (temperatures from $\sim$1 to at least 10 MK ) using much finer resolution ($\sim$1 arcsec/pixel) than is observed with the various instruments used by these authors (i.e. LASCO (temperature-insensitive white light; C2: $\sim$11.4 arcsec/pix), EIT ($\sim$0.6-3 MK; $\sim$2.6 arcsec/pix), UVCS ($\sim$2-8 MK; $>$ $\sim$40 arcsec/bin for high temperatures)).  

\begin{figure}[!ht] 
\begin{center}
\includegraphics[width=0.8\textwidth]{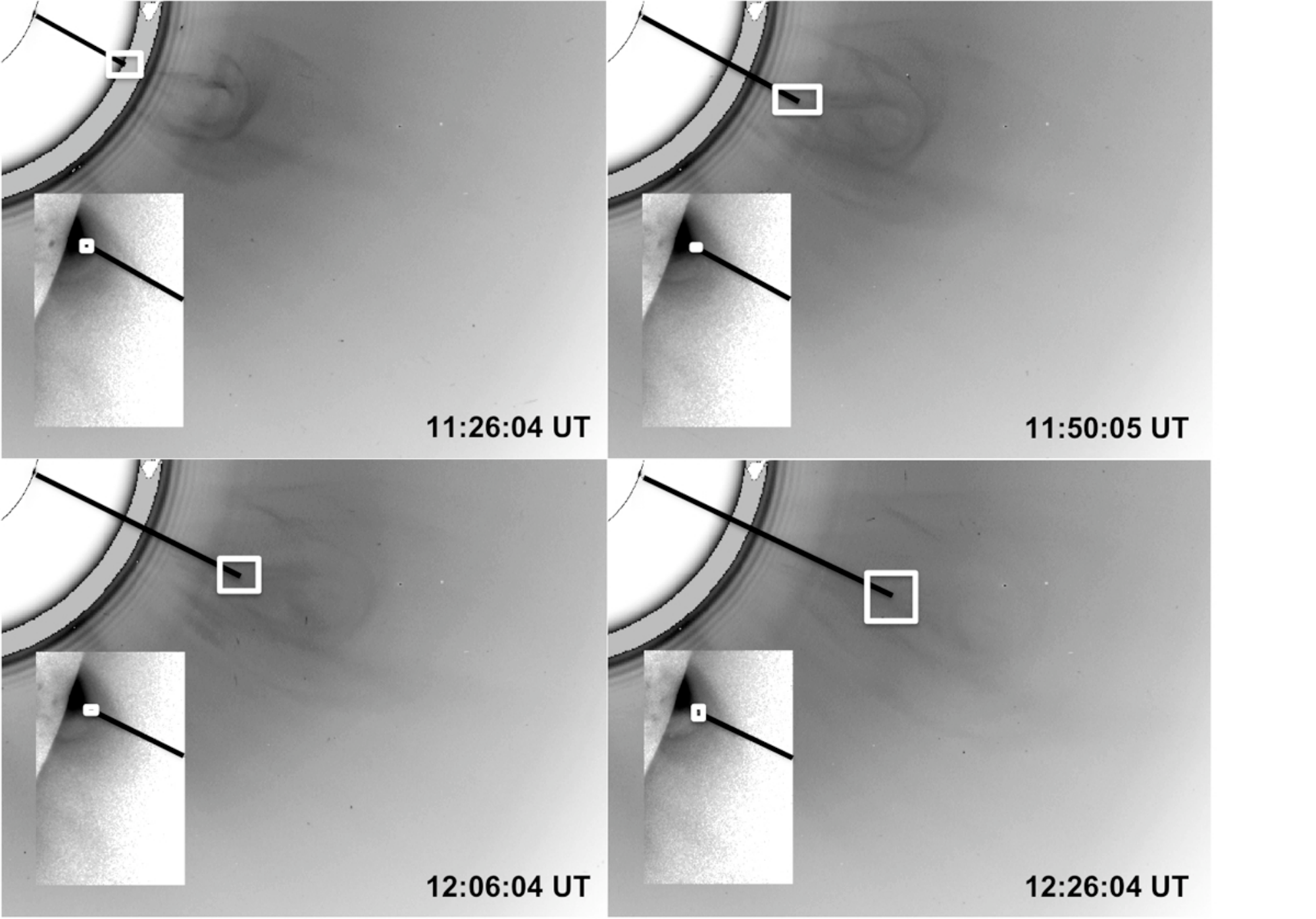}
\caption{Reverse-scale composite images with an enlarged XRT inset overlaid onto corresponding LASCO C2 images.  The length of the candidate current sheet is represented by the full solid line within the LASCO FOV.  The ``p" and ``q" values correspond to the center of the white error boxes in the XRT and LASCO images, respectively.  The times indicated are those of the LASCO images.  The corresponding times for the inset XRT images differ by less than one minute.}
\label{cs_qp_images}
\end{center}
\end{figure}

Because both XRT and LASCO observed this event, it is also possible to measure the length of the CCS assuming that the hot CCS observed in the XRT images corresponds to the white-light features in the LASCO images.  Figure~\ref{cs_qp_images} depicts the progression of the CME height.  The full length of the CCS is displayed as the solid dark line in the LASCO FOV.  An enlarged XRT FOV is inset for each LASCO image to show the position of the CCS just above the XRT arcade.   The white boxes represent the error manually assigned to each position.  The center of the boxes in the XRT images corresponds to the ``p" value as described in \cite{linforbes00} and \cite{webb03} as the bottom position of the current sheet where it meets the top of the arcade.  
Conversely, the LASCO box indicates the ``q" value which is the top position of the current sheet where it meets the bottom of the CME.  (See Figure~\ref{cs_cartoon} for a visual reference.)  These positions, determined ``by eye", have been de-projected based upon the footpoint used to de-project the flows (see Figure~\ref{tracks}).  It is important to note that these current sheet lengths are assumed to be the distance between these ``p" and ``q" values rather than direct length measurements of a confirmed current sheet; therefore, these reported lengths are upper bounds.

\begin{figure}[!ht] 
\begin{center}
\includegraphics[width=0.6\textwidth]{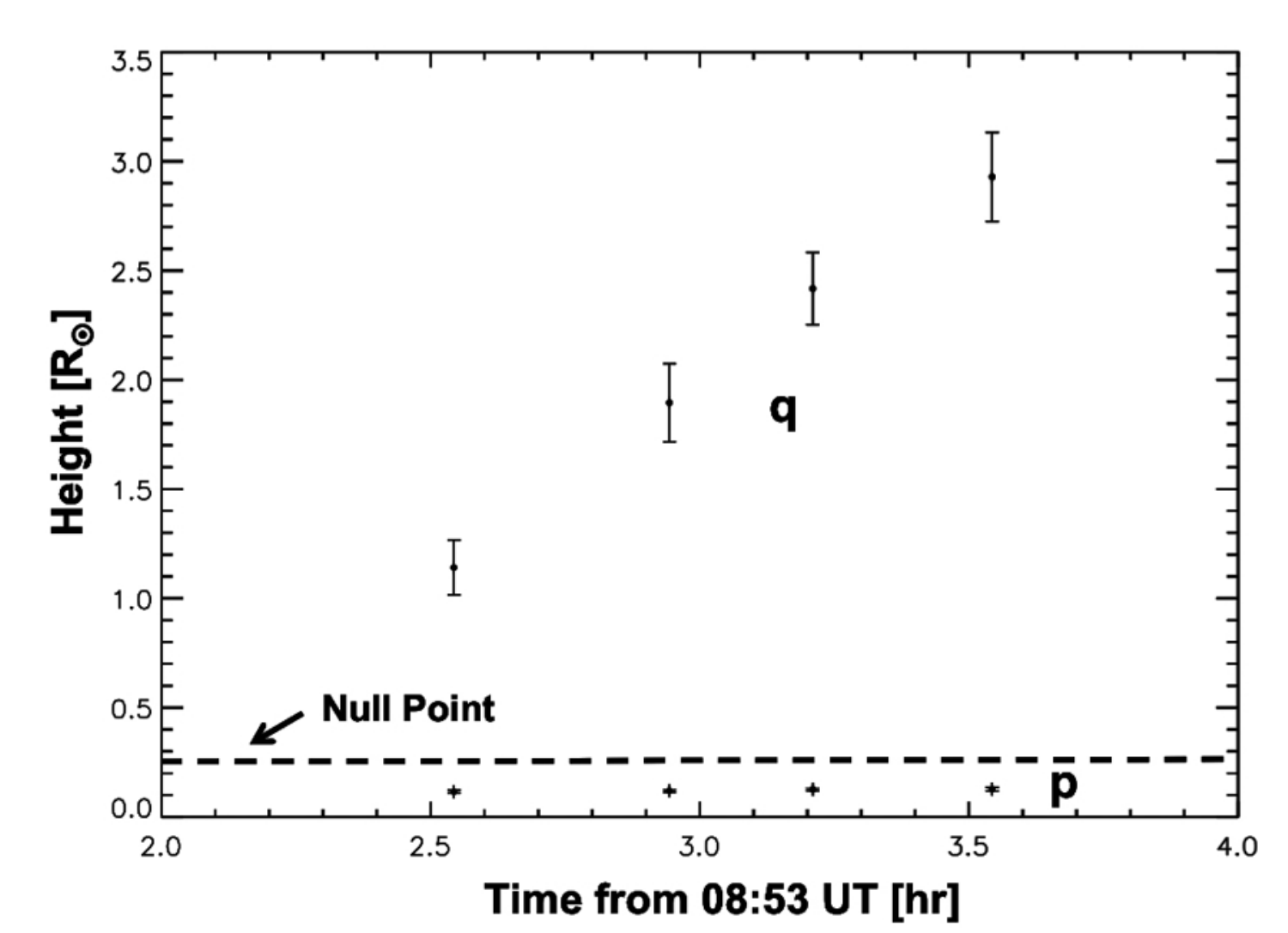}
\caption{De-projected ``q" and ``p" heights in solar radii versus the time elapsed since the loss of equilibrium as seen in the \textit{STEREO A}/SECCHI data (approximately 08:53 UT).  (Refer to Figure~\ref{cs_cartoon} for a qualitative reference of the ``q" and ``p" positions.)  This plot qualitatively agrees with predictions from the \cite{linforbes00} reconnection model.  (See Figure 9 from \cite{webb03} for a comparative graph.)  The dashed horizontal line indicates the height of the null point (0.25 R$_{\odot}$ $\sim$ 170 Mm above the surface) which is discussed in Section 4.}
\label{cs_qp_plot}
\end{center}
\end{figure}

The results of the CCS length measurements are shown in Figure~\ref{cs_qp_plot}.  There is strong qualitative agreement with the predictions from the \cite{linforbes00} reconnection model.  A useful composite plot for comparing Figure~\ref{cs_qp_plot} with the \citeauthor{linforbes00} height versus time predictions is given by \cite{webb03} (see Figure 9 therein).

\clearpage

The CCS length measurements correspond to an average thickness (as measured within the XRT FOV) of $\sim$4.5 Mm.  The de-projected length measured at the time of the first and last CME detections in the LASCO C2 FOV is $\sim$710 and 1950 Mm, respectively.  The Alfv\'{e}n Mach number can be calculated as M$_{A}$ = thickness / (q - p) resulting in estimates of 0.006 - 0.002 (in order of LASCO C2 CME detection).  These values are lower than previous estimates for other flares of greater than 0.1 (\citeauthor{ciaray08} \citeyear{ciaray08}; \citeauthor{linetal05} \citeyear{linetal05}; \citeauthor{webb03} \citeyear{webb03}; \citeauthor{ciaetal02} \citeyear{ciaetal02}; \citeauthor{ko02} \citeyear{ko02}) and rival the prediction by \cite{linforbes00} requiring an M$_{A}$ of at least 0.005.  This prediction, however, applies to high-speed CMEs ($>$ 1000 km s$^{-1}$).  The average speed of the 2008 April 9 CME within the LASCO C2 FOV is only about 450 km s$^{-1}$ \citep{landi10}; therefore, the Mach number requirements for this model of reconnection may be relaxed.  The Mach number predictions also vary greatly depending on the model's inclusion of 3-D geometry.

\subsection{Shrinking Loops \& Flows} 

\begin{figure}[!ht] 
\begin{center}
\includegraphics[width=1.\textwidth]{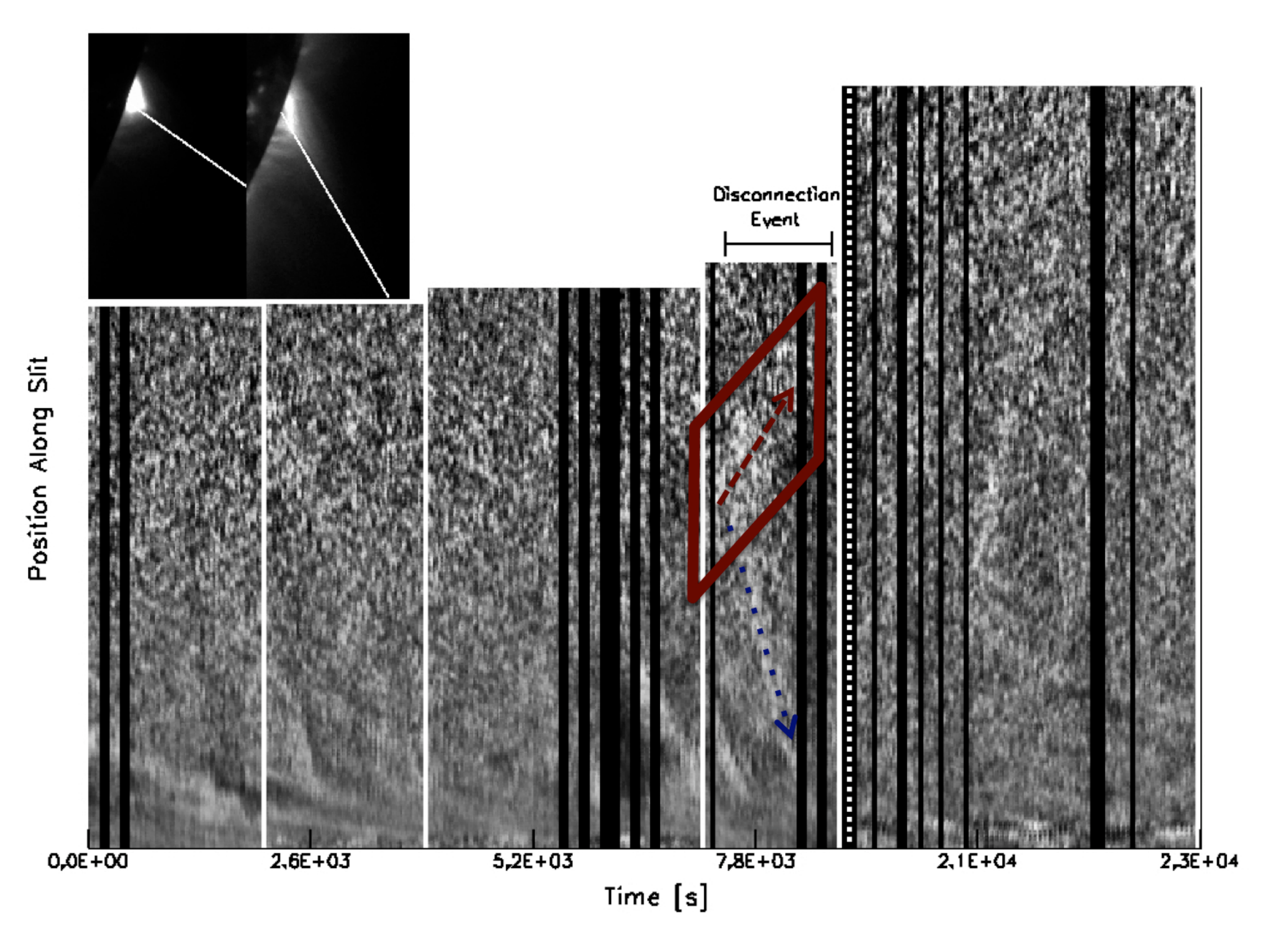}
\caption{Position versus time stackplot created with three-pixel wide slits taken along the current sheet candidate as it progressed southward within the XRT FOV.  Representative slits are displayed as insets in the top-lefthand corner.  Each three-pixel segment represents approximately one minute.  Dark and bright lanes indicate motion along the extractions.  The black vertical strips are due to small data gaps, and the dashed vertical strip represents the large data gap beginning at 16:25 UT.  The separate panels indicate that a different slit was used for that time frame to allow for the southward progression.  The disassociation position of the reconnection outflow pair is indicated in the figure.  The upflow portion of the disconnection event is much more diffuse than its downflow counterpart, making it appear broader and dimmer.}
\label{stack_plot}
\end{center}
\end{figure}

Following the CME and beginning at approximately 11 UT, several individual, dark shrinking loops can be seen retracting just above the arcade.  Bright regions can be seen flowing into the apex of the arcade beginning at 12:54 UT.  (Dark and bright loops retracting just above the arcade are also clearly observed by SECCHI in the 171{\AA}  and 284{\AA}  bandpasses.)  These bright regions are broader than the dark shrinking loops seen earlier and are more difficult to distinguish as individual loops.  Loop structure associated with these regions is still maintained, however, in some of the images through which they track.  Sharpening the data also reveals flows moving outwards from the flaring region.  A three-pixel wide extraction is taken from each image along the CCS to create the stackplot image shown in Figure~\ref{stack_plot}.  The dark and bright lanes in the stackplot are due to the motion of the loops and flows along the CCS.  (Hereafter, the apexes of the shrinking loops will also be referred to as downflows.)

A particularly intriguing flow set occurs between 12:18 UT and 12:35 UT and is labeled as the ``Disconnection Event" in Figure~\ref{stack_plot}.  A bright region appears, moves slightly towards the arcade, and then disassociates.  Part of the region flows towards the arcade while the other flows in the opposite direction away from the arcade along the CCS.  To our knowledge, a possible reconnection outflow pair episode such as this (i.e. appearing in a region where retracting loops and upflows have been observed and along a directly observable SXR current sheet) has not been observed this close to the solar surface (at nearly 190 Mm).  Outflow pairs much higher in the corona have been previously observed primarily with LASCO (see for example \citeauthor{sheeley07} \citeyear{sheeley07}, \citeauthor{linetal05} \citeyear{linetal05}, and \citeauthor{sheeley04} \citeyear{sheeley04}).  Possibly related features have been observed with RHESSI (e.g. \citeauthor{liu08} \citeyear{liu08}; \citeauthor{sui04} \citeyear{sui04}; \citeauthor{sui03} \citeyear{sui03}) and various other instruments ranging from H-alpha to SXR detectors along with white-light coronagraphs (e.g. \citeauthor{trip07} \citeyear{trip07}; \citeauthor{trip06_a} \citeyear{trip06_a}), though the interpretations may differ.  The large data gap unfortunately occurs immediately following this event hindering possible observations of continued reconnection occurring in the region.  

Previous work done by the authors has focused on automatically tracking and characterizing supra-arcade downflows (SADs) during long duration flaring events \citep{sadsI}.  Automatically tracking shrinking loops for this event presents several challenges primarily due to the low signal to noise ratio complicated by the southward progression of the CCS.  Therefore, the flows from this flare are tracked manually to ensure reliability.    

\begin{figure}[!ht] 
\begin{center}
\includegraphics[width=.8\textwidth]{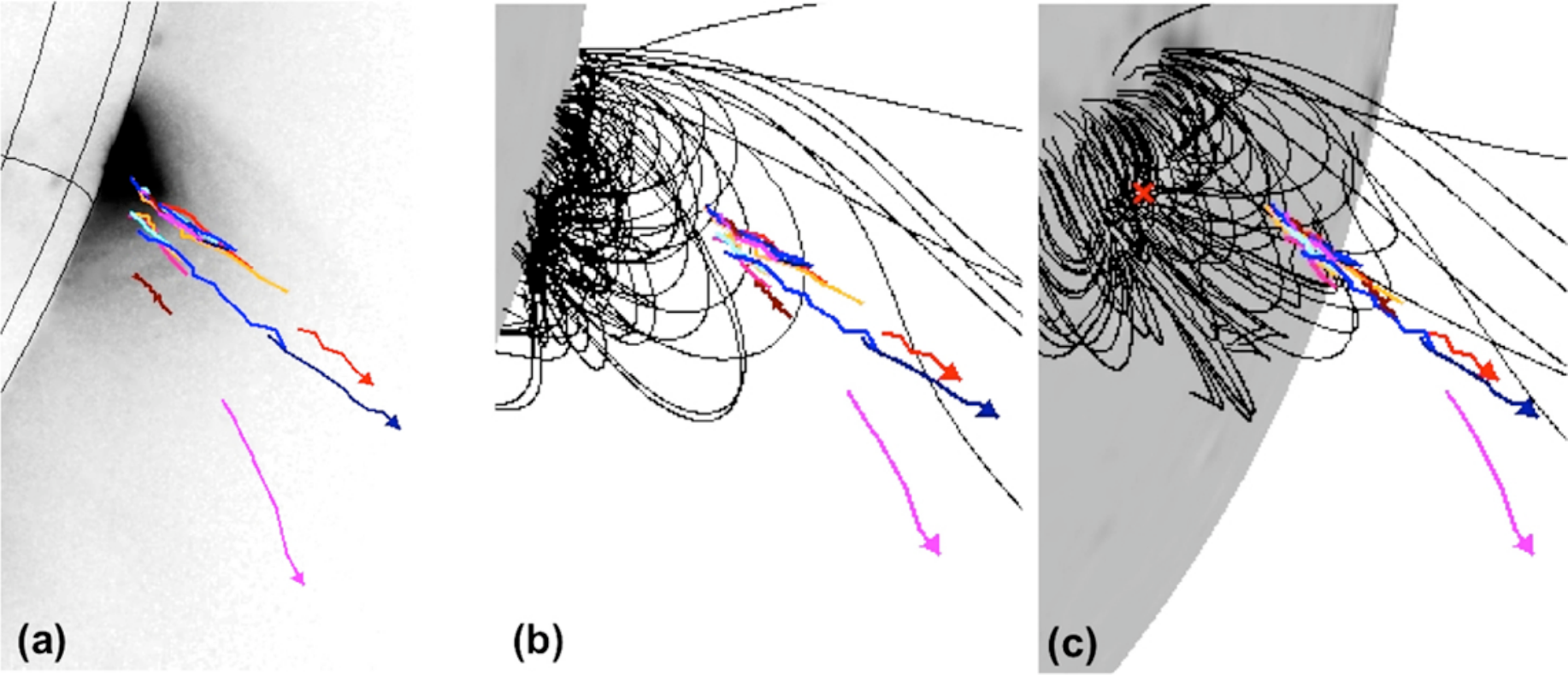}
\caption{(a) Plane-of-sky trajectories of the manually-tracked flows overlaid onto a reverse-scaled XRT image taken at 16:30:54 UT.  (b) \& (c) De-projected trajectories based on a convergent footpoint at a latitude of $-$18$^{\circ}$ and a longitude of 23$^{\circ}$ beyond the west limb.  The reverse-scaled background images (MDI magnetogram plus magnetic loops) are derived from a PFSS model package \citep{pfss03} with a center longitude set to (b) 115$^{\circ}$ and (c) 150$^{\circ}$.   The flows are assumed to have no velocity component in the longitudinal direction.  The arrows represent upflow trajectories.}
\label{tracks}
\end{center}
\end{figure}

A total of sixteen flows are tracked (13 downflows and 3 upflows) over a total time of 6.22 hours between 11:11 and 17:24 UT.  The number of frames through which each flow is observed ranges from 3 to 22 with a median frame count of 10.  Figure~\ref{tracks} (a) shows the plane-of-sky trajectories of the manually-tracked flows.  Note that the magenta flow in the southwest region represents the track of a very faint, diffuse upflow that occurred after the large data gap.  Despite its discrepant path compared to the other flows, it still tracks along the CCS which had progressed southward by the time of this flow.

We use a Potential Field Source Surface (PFSS, \citeauthor{pfss03} \citeyear{pfss03}) model to extrapolate the magnetic field from active region 10989.  The magnetic field morphology suggests that the footpoints of the shrinking loops lie near a latitude of $-$18$^{\circ}$ and a longitude of 23$^{\circ}$ beyond the west limb.  This footpoint is used as a point of convergence for all of the flows in order to de-project their positions above the solar surface.  The flows are assumed to have no velocity component in the longitudinal direction.  The positions have also been rotated based on the latitude of the footpoint.  The resulting de-projected trajectories are shown in Figure~\ref{tracks} (b) and (c) overlaid onto reverse-scaled MDI magnetograms with magnetic loops traced out from the PFSS modeling.  We emphasize that 2-D motion is measured from the images, in plane of the sky.  The uncertainty in the 3-D trajectory is entirely due to assumed longitude.

\clearpage

\begin{figure}[!ht] 
\begin{center}
\includegraphics[width=0.45\textwidth]{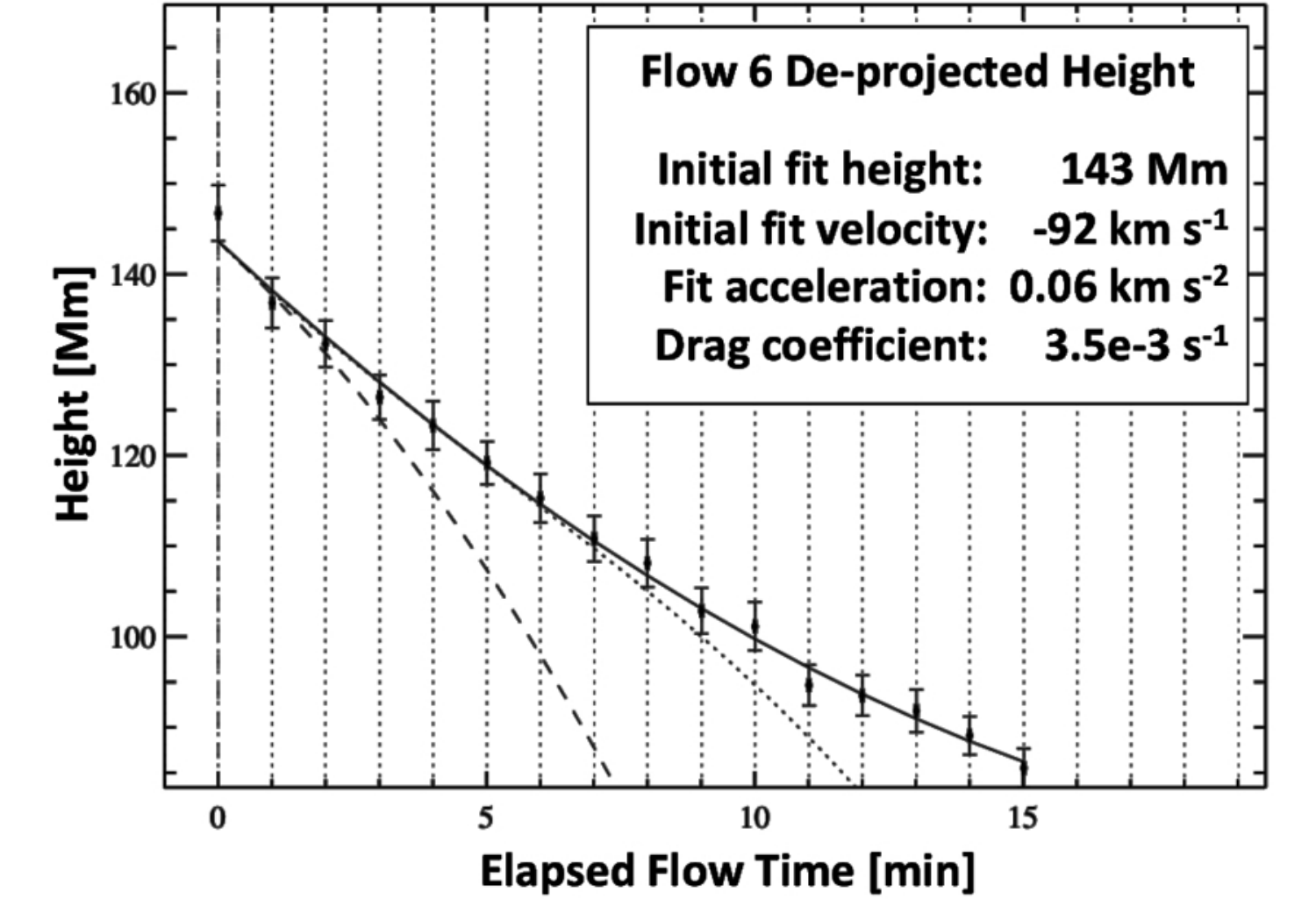}
\includegraphics[width=0.45\textwidth]{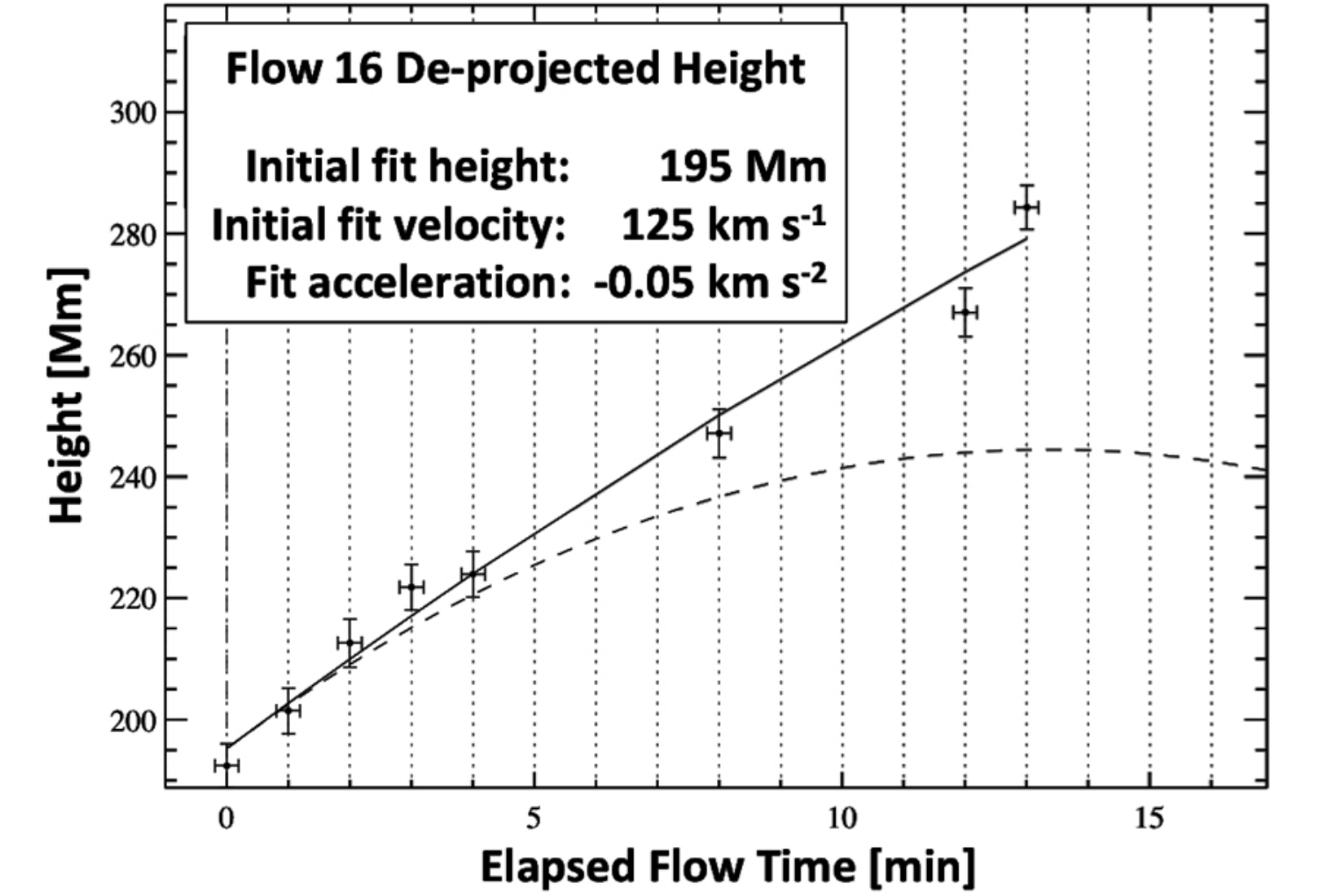}
\caption{De-projected height-time profiles represented by (a) a typical downflow and (b) a typical upflow observed in this study.  The vertical dotted lines mark the mid-exposure time of each image in the sequence.  (Note that Flow 16 does not have a contiguously-detected track due to the low signal-to-noise ratio at the heights through which it travels.)  The solid line is the 2-D polynomial fit applied to the profile to obtain the initial velocity and acceleration.  The calculated fit parameters are given in the legend for each flow.  The thick dashed line represents the gravitational profile for a body in free-fall given the initial height and velocity of the flow.  The thin dashed profile line in (a) represents the gravitational profile for a body in free-fall experiencing a constant drag coefficient of 3.5x10$^{-3}$ s$^{-1}$.}
\label{flow_profiles}
\end{center}
\end{figure}

Figure~\ref{flow_profiles} shows de-projected height-time profiles for (a) a typical downflow and (b) a typical upflow observed in this study.  The positions are given with respect to the chosen footpoint on the solar surface.  The error bars displayed for the position represent the square root of the size of the flow at that position.  These extents are determined manually and are chosen conservatively; however, errors associated with footpoint assignment and tilt in the longitudinal direction are not included in the error estimate.  The initial velocity and acceleration uncertainties are determined by adding these position errors to the 2-D polynomial fit.  This fit is represented by the solid profile line in Figure~\ref{flow_profiles}.  The gravitational profile for a body in free-fall given the initial height and velocity of the flow is shown as the thick dashed line.  The left panel includes a thin dashed profile line outlining the trajectory of a body in free-fall experiencing a constant drag force.

\begin{figure}[!ht] 
\begin{center}
\includegraphics[width=.6\textwidth]{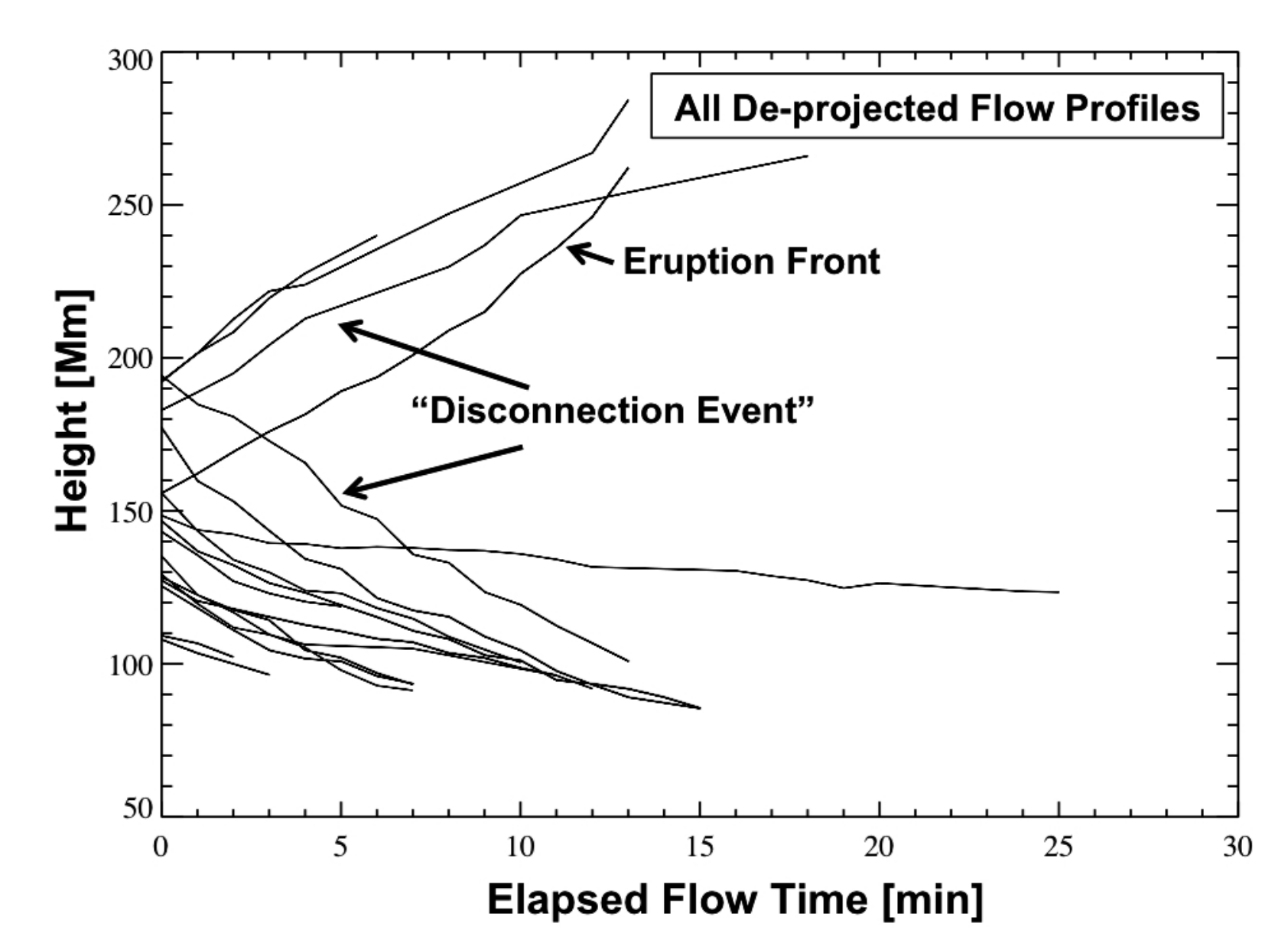}
\caption{All de-projected height-time profiles.  Profiles with a positive slope represent upflows.  The eruption front and disconnection event profiles are labeled in the figure.  Note that the initial heights of the upflow and downflow portions of the disconnection event are separated slightly because the region appears to brighten and move sunward prior to separating.}
\label{all_flow_profiles}
\end{center}
\end{figure}

All of the downflow speeds are slower than their corresponding free-fall speeds especially as they near the limb, except for one flow that is only tracked through three frames.  Faster downflow speeds would have supported the reconnection outflow hypothesis; however, it is not excluded by this opposite result either considering that other flow characteristics provide significant support.  Namely, several of the flows exhibit a clearly-defined cusped or rounded loop structure and one of the regions is observed to disconnect into an upflow and a downflow which are expected results from reconnection.  Also, the flow profiles diverge the most from the gravitational profiles as they near the limb and, presumably, the top of the arcade where the loops are expected to settle to a potential configuration.  If these flows are indeed reconnection outflows, then this result may indicate that a source of drag (e.g. mass build up in front of the flows due to density in the current sheet, shocks, magnetic field entanglement, etc.) has a significant effect on the flow speeds.  Additionally, these profiles are qualitatively supported by reconnection models.  \cite{lin04} (Figure 5 therein) has shown that the retracting reconnected loops shrink primarily within the first 10-20 minutes and then decelerate considerably which is consistent with the profiles shown in Figure~\ref{all_flow_profiles}.

The time for each position shown in Figure~\ref{flow_profiles} is taken as the frame start time plus half of the exposure duration; thus, the time error bars, too small to be visible on the plots, represent half of the exposure duration which ranges from 2 to 16 seconds.  The vertical dotted lines mark the time of each image (start time plus half of the exposure duration) in the sequence.  Note that due to a low signal to noise ratio high above the arcade, the flows are not always detected in contiguous frames.  The gaps in the profile in Figure~\ref{flow_profiles} (b) (i.e. vertical dashed lines without a flow position marked) graphically represent this discontinuity.  A 2-D polynomial fit (solid line) is applied to each profile to obtain initial velocities and accelerations.  All of the profiles are plotted together in Figure~\ref{all_flow_profiles}.  

\clearpage

Because expected quantitative predictions can vary largely within reconnection models depending on conditions and constraints, a qualitative way of distinguishing between the Sweet-Parker and Petschek models is through the profiles of the flow trajectories.  Sweet-Parker predicts a sustained acceleration all along the current sheet with the Alfv\'{e}n speed being reached only near the tips of the sheet.  The Petschek model predicts acceleration in a small diffusion region near the reconnection point with the flow's final velocity (nominally the Alfv\'{e}n speed) being reached within a relatively short distance near the reconnection point and then maintained until it exits the current layer (including slow shocks).  The measured flow profiles exhibit accelerations very near to 0 km s$^{-2}$, except for some slowing as the downflows approach the post-eruption arcade, which matches well with the Petschek predictions \citep{somov92}.

\begin{figure}[!ht] 
\begin{center}
\includegraphics[width=.49\textwidth]{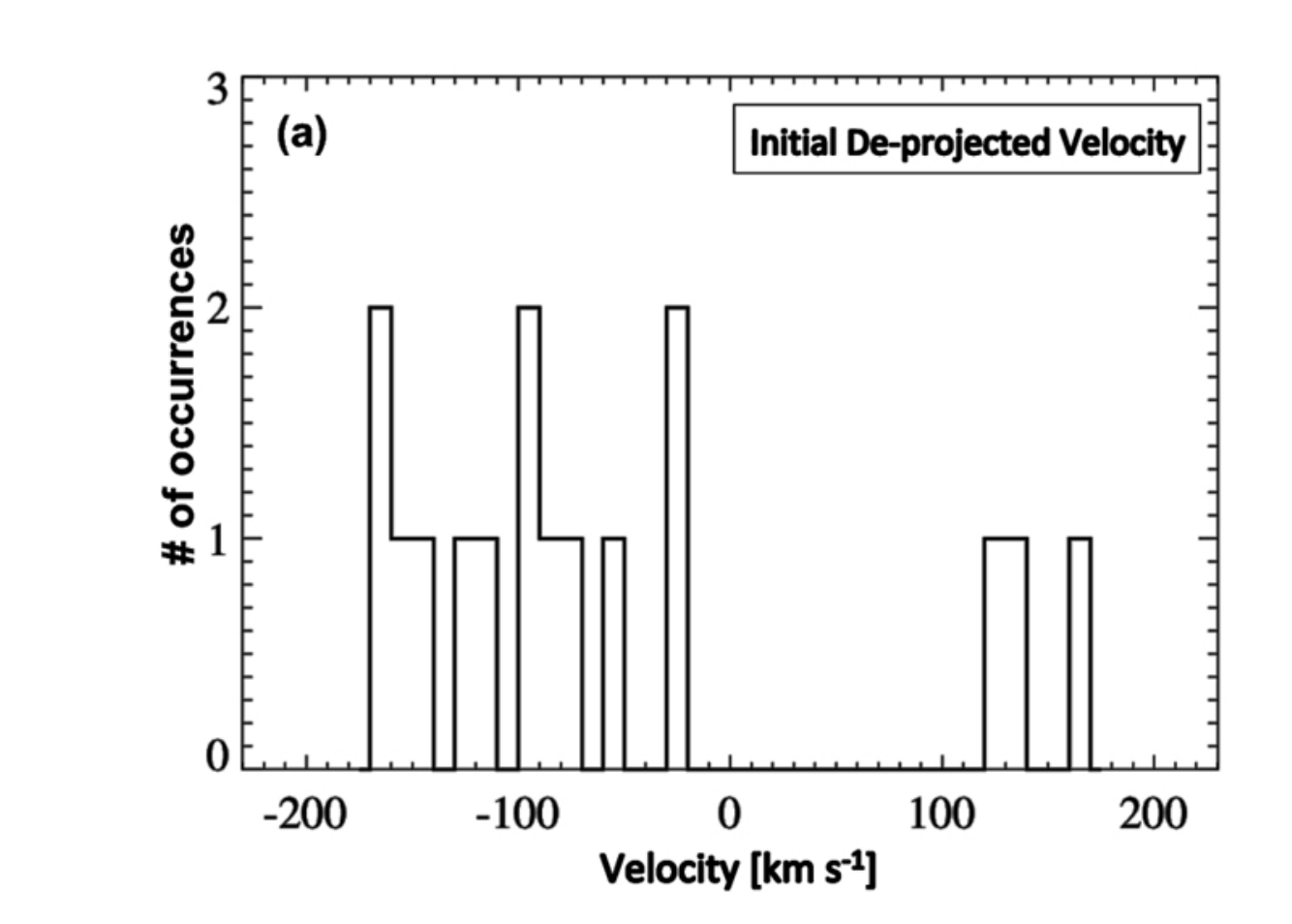}
\includegraphics[width=.49\textwidth]{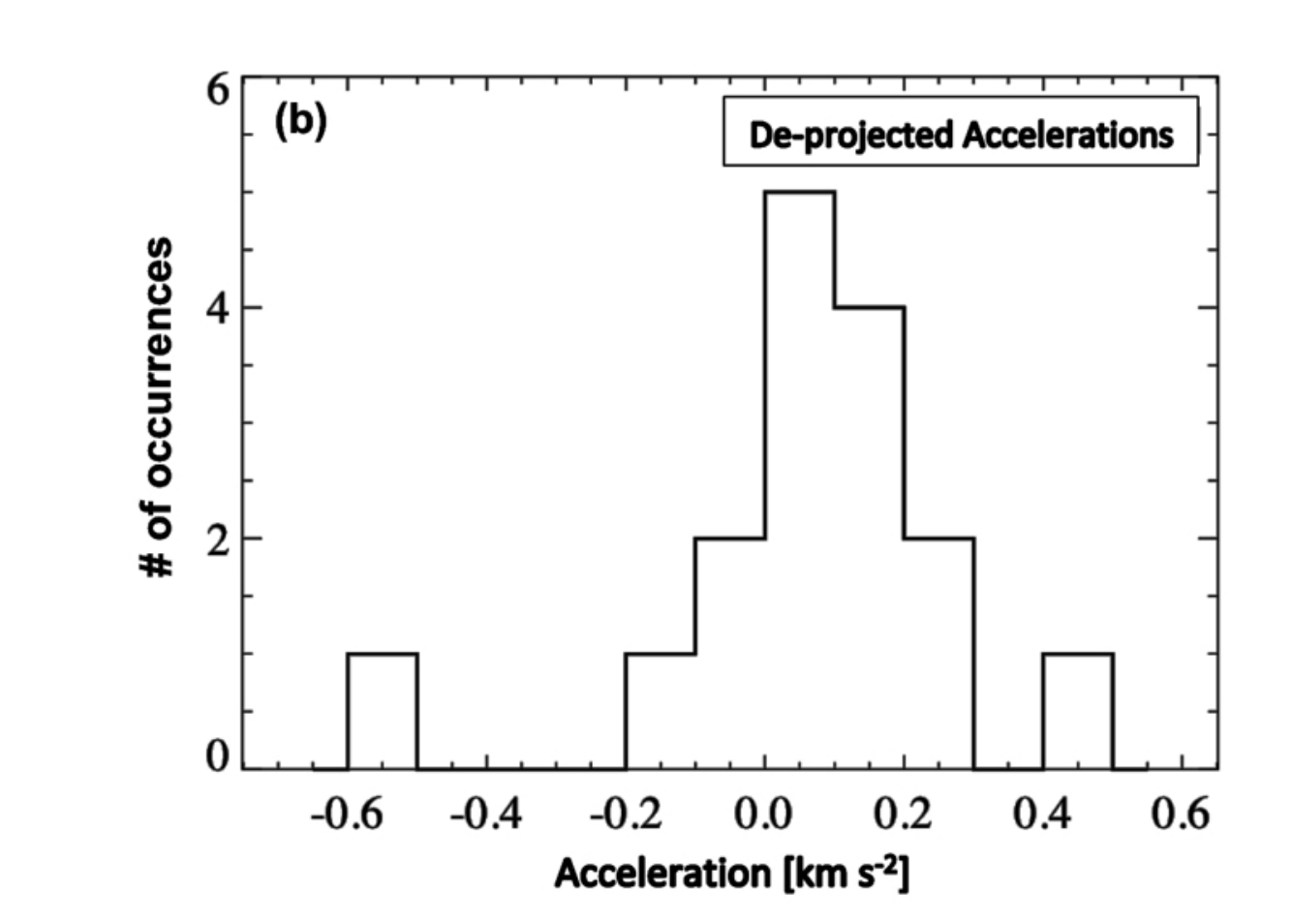}
\caption{(a)  Histogram of the initial de-projected velocities.  Upflow velocities are positive.  (b)  Histogram of the de-projected accelerations.  All flows are slowing down or have accelerations consistent with zero.}
\label{hist_va}
\end{center}
\end{figure}

The de-projected speeds vary from the plane-of-sky speeds by up to 11\% but with a mean difference of only 7\%.  The initial de-projected speeds range from 21 to 165 km s$^{-1}$ (median:  120; mean:  109).  The slowest speed is attributed to the last shrinking loop occurring after the large data gap.  See Figure~\ref{hist_va} (a) for a histogram of these velocities.  Upflows have been assigned a positive value.  The distribution from this small sample does not yield a recognizable trend; however, all speeds are much smaller than the predicted value near the Alfv\'{e}n speed which is typically assumed to be a value on order of 1000 km s$^{-1}$ for the corona.  We are unable to directly measure the coronal magnetic field and thus, the Alfv\'{e}n speed.

The de-projected acceleration distribution is displayed in Figure~\ref{hist_va} (b).  The acceleration magnitudes range from 0.01 to 0.6 km s$^{-2}$ (median:  0.1; mean:  0.2).   All flows are either slowing down slightly or have average accelerations consistent with zero.

\begin{figure}[!ht] 
\begin{center}
\includegraphics[width=.49\textwidth]{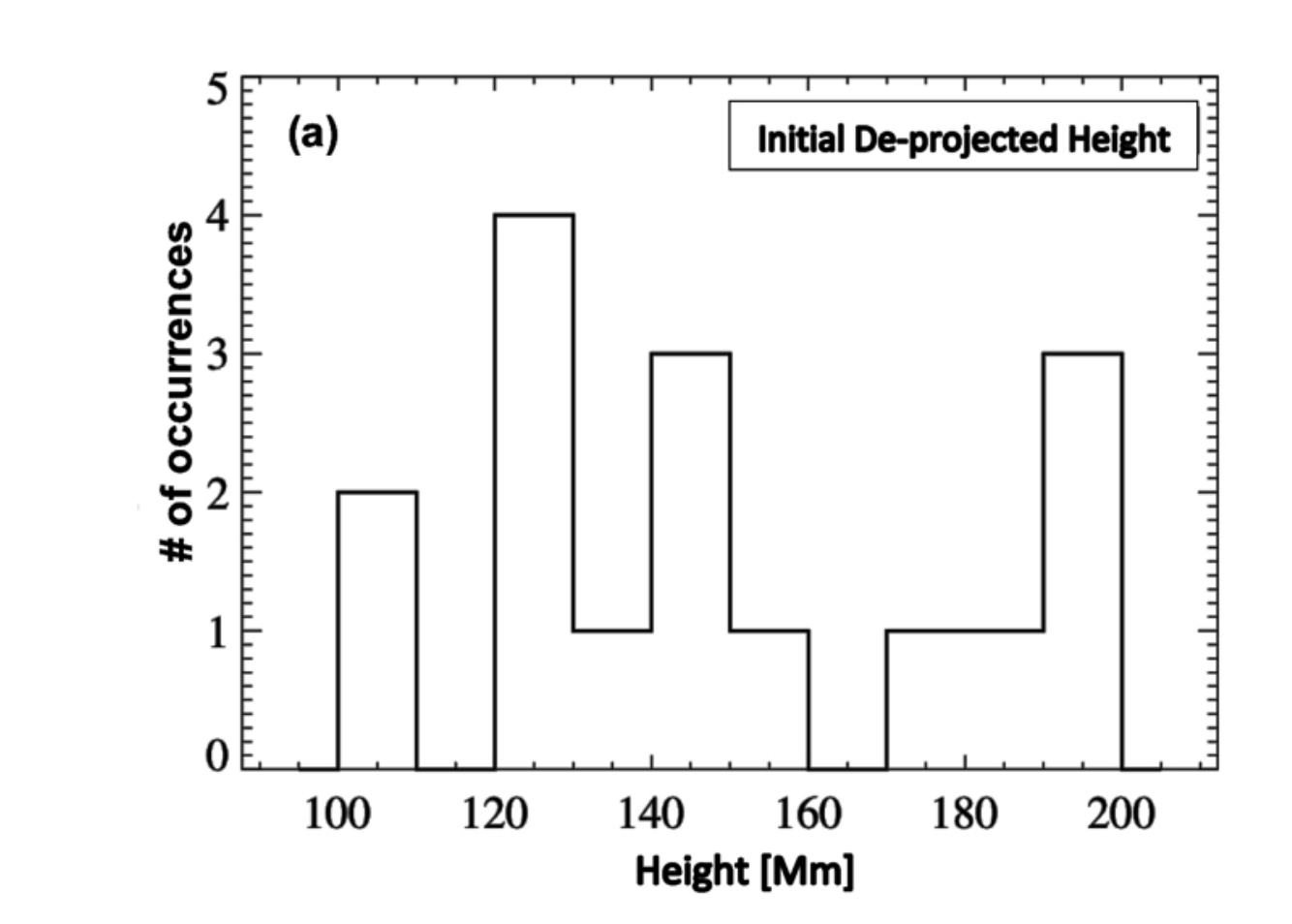}
\includegraphics[width=.49\textwidth]{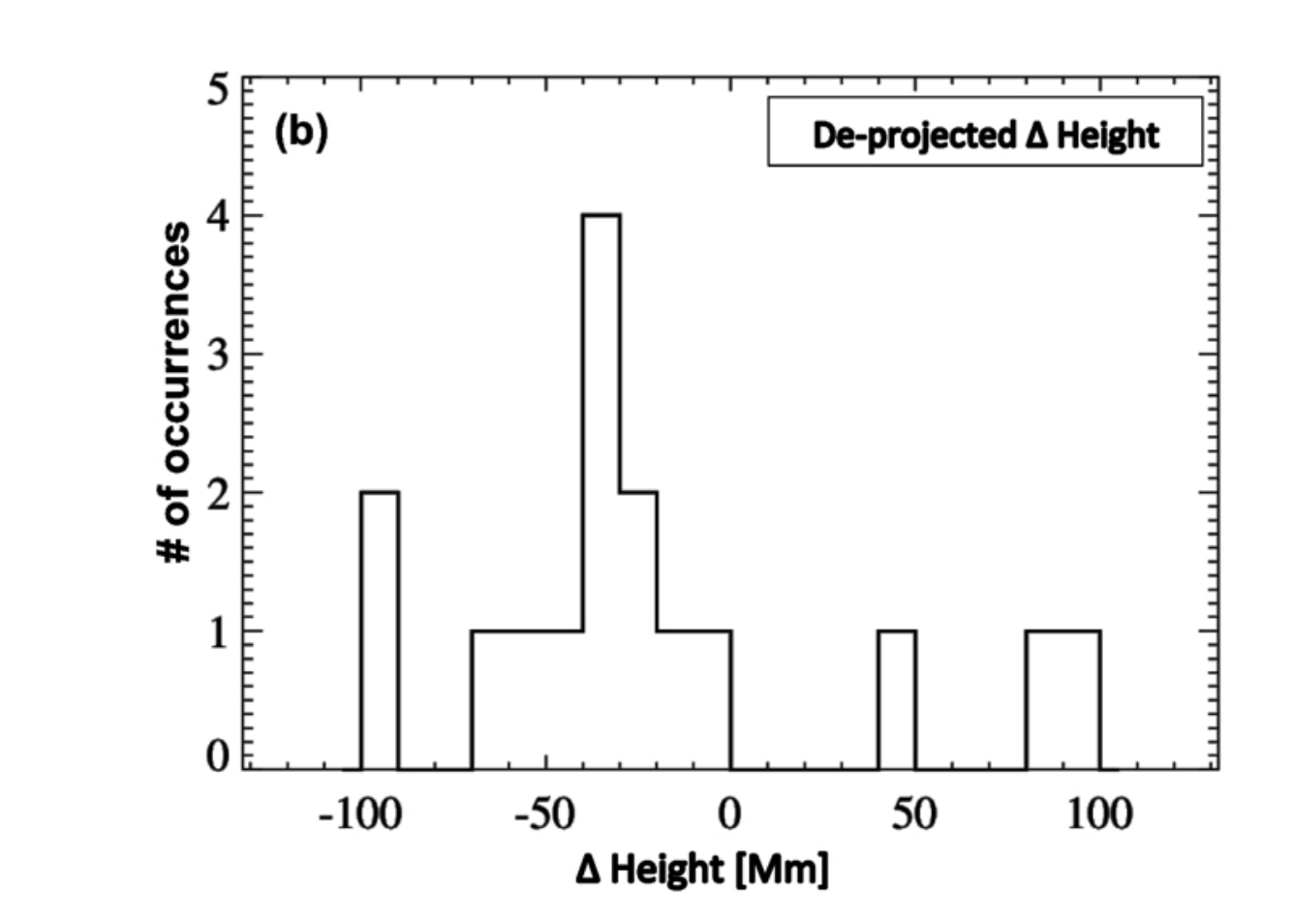}
\caption{(a)  Histogram of the initial de-projected heights.  (b)  Histogram of the de-projected change in heights.}
\label{hist_h}
\end{center}
\end{figure}

\clearpage

The initial de-projected heights above the footpoint, which vary from the plane-of-sky heights by up to 10\%, are displayed in Figure~\ref{hist_h} (a).  The heights range from 108 to 194 Mm (median:  145; mean:  150).  The overall change in de-projected height (i.e. ``distance travelled") of the flows is shown in Figure~\ref{hist_h} (b).  These values range from 7 to 106 Mm (median:  40; mean:  48).  Previously-measured downflow estimates for similar limb flares have upper limits of about 130 Mm on height and about 40 Mm on the shrinkage distance \citep{sadsI}.  The increase in these height measurements for the ``Cartwheel CME" flare can be attributed to the footpoints being occulted (as discussed in Section 2).

\subsection{XRT \& LASCO Upflows}

Further inquiry was performed to determine how the upflows tracked beyond the XRT FOV.  The upflows are detected in the noisiest regime of the images high above the flare arcade and are therefore the most difficult to accurately track.  Combining these relatively high-error position estimates with fewer detected positions due to their location towards the edge of the FOV adds a large uncertainty to extrapolating their trajectories beyond the XRT FOV.  

Despite these difficulties, extrapolating the upflow positions simply using their average speeds revealed some possible associations with flows observed in LASCO.  In an attempt to incorporate a more accurate profile for the XRT upflows into the outer corona, the corresponding LASCO flows are tracked and an acceleration is determined from their height-time profiles.  Those accelerations are determined to be 0.025, 0.02, and 0 km s$^{-2}$ respectively.  (The path of the third LASCO flow is difficult to track and yielded inconsistent results; therefore, an acceleration of 0 km s$^{-2}$ is applied.)

\begin{figure}[!ht] 
\begin{center}
\includegraphics[width=1\textwidth]{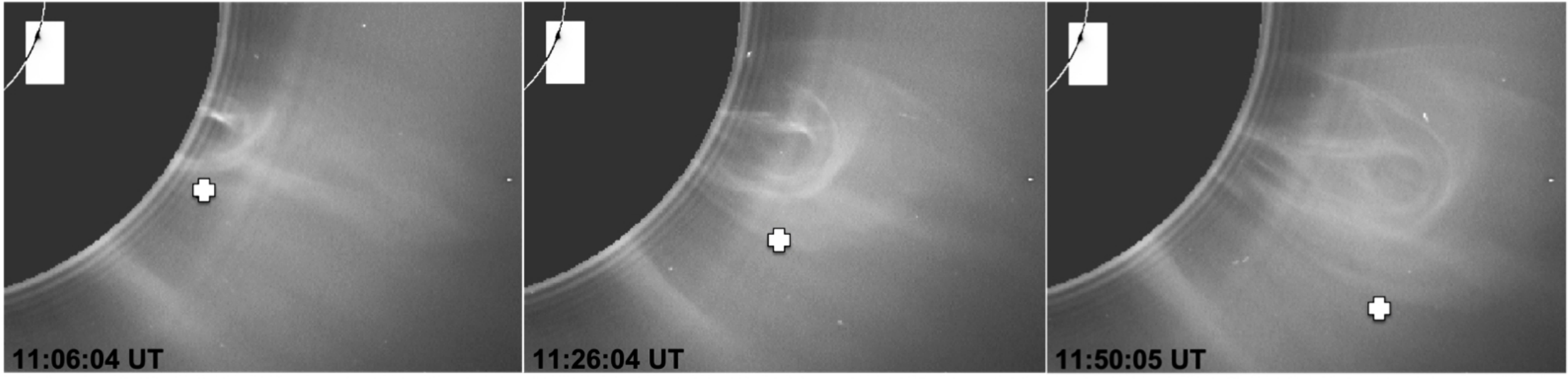}
\includegraphics[width=1\textwidth]{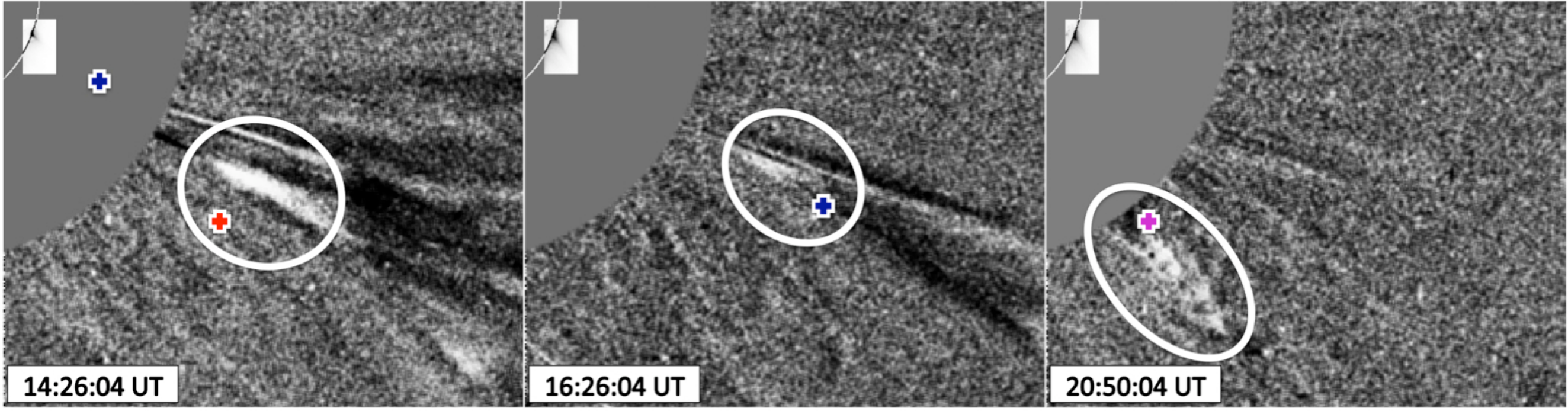}
\caption{\textbf{XRT \& LASCO flow associations}:  All panels consist of composite XRT (reverse-scaled)/LASCO C2 images.  \textbf{\textit{Top Panel}:}  The white crosses correspond to the extrapolated position of the XRT eruption front which tracks precisely with the white-light CME seen in the LASCO images.  The acceleration of the LASCO CME is factored into the extrapolations; however, deflection is not included (see Figure~\ref{cme_path}).  \textbf{\textit{Bottom Panel}:}  The LASCO images are running-mean-differenced and enhanced to emphasize flows. The colored crosses correspond to the extrapolated positions of the XRT upflows.  Ellipses indicate strong XRT/LASCO flow associations after using the LASCO flow accelerations to extrapolate the XRT upflow positions.  Note that the colors in this figure correspond to the colors of the tracks in Figure~\ref{tracks}.}
\label{flow_asscn}
\end{center}
\end{figure}

The XRT upflow positions are extrapolated to the outer corona by using their final fit velocity in the XRT FOV as an initial velocity at the edge of the LASCO FOV.  The corresponding LASCO flow accelerations are then applied to determine successive flow positions.  The XRT upflow paths are assumed to be straight although the CME path itself initially veers northward (see Figure~\ref{cme_path}).  This results in the upflows tracking just to the south of the LASCO flows.  As a check on this procedure, the eruption front position, which is expected to be observed in LASCO as a white-light CME, is extrapolated in the same manner using a measured acceleration of 0.03 km s$^{-2}$.  Allowing for some angular separation due to the aforementioned CME path deflection, the resulting extrapolated positions correspond precisely to the CME front in the LASCO FOV (Figure~\ref{flow_asscn}, top panel).

\clearpage

This minimally-biased method results in strikingly close associations between the XRT and LASCO upflows (Figure~\ref{flow_asscn}, bottom panel).  In addition, the upflow resulting from the XRT disconnection event appears to closely correspond with the concave upward feature following the CME (i.e. the CME ``pinch-off point") which in turn corresponds to the CME disconnection event described by \cite{webb03}.

\clearpage

\section{Discussion}

Some standard eruptive flare models consist of a flux rope which is released as a coronal mass ejection, generally due to some reconnection occurring near the active region.  The flux rope eruption is followed by the formation of a current sheet between the CME and the growing underlying arcade along the polarity inversion line.  The arcade is formed when magnetic loops reconnect high in the current sheet and then retract to a less energetic configuration.

\begin{figure}[!ht] 
\begin{center}
\includegraphics[width=.45\textwidth]{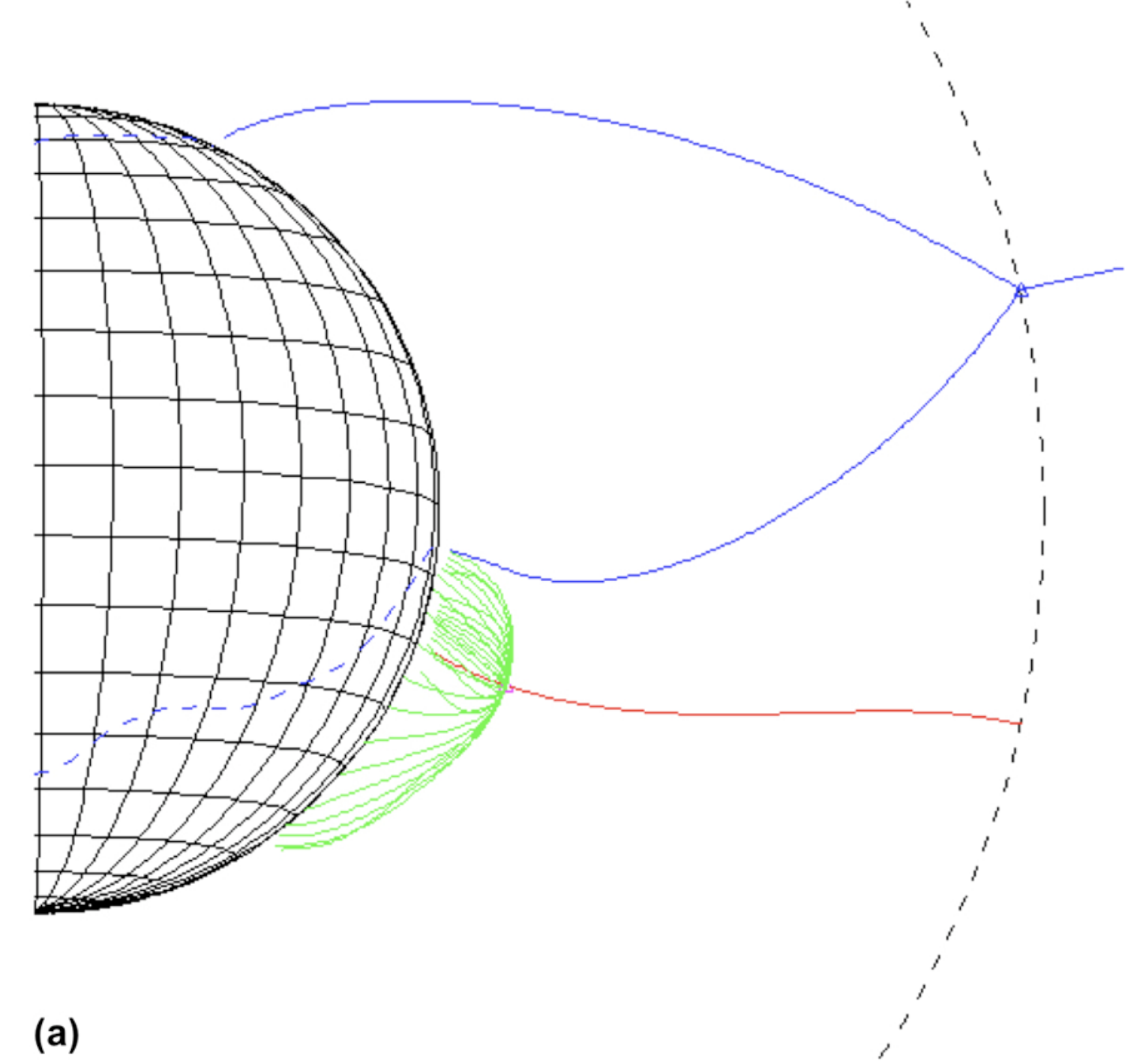}
\includegraphics[width=.5\textwidth]{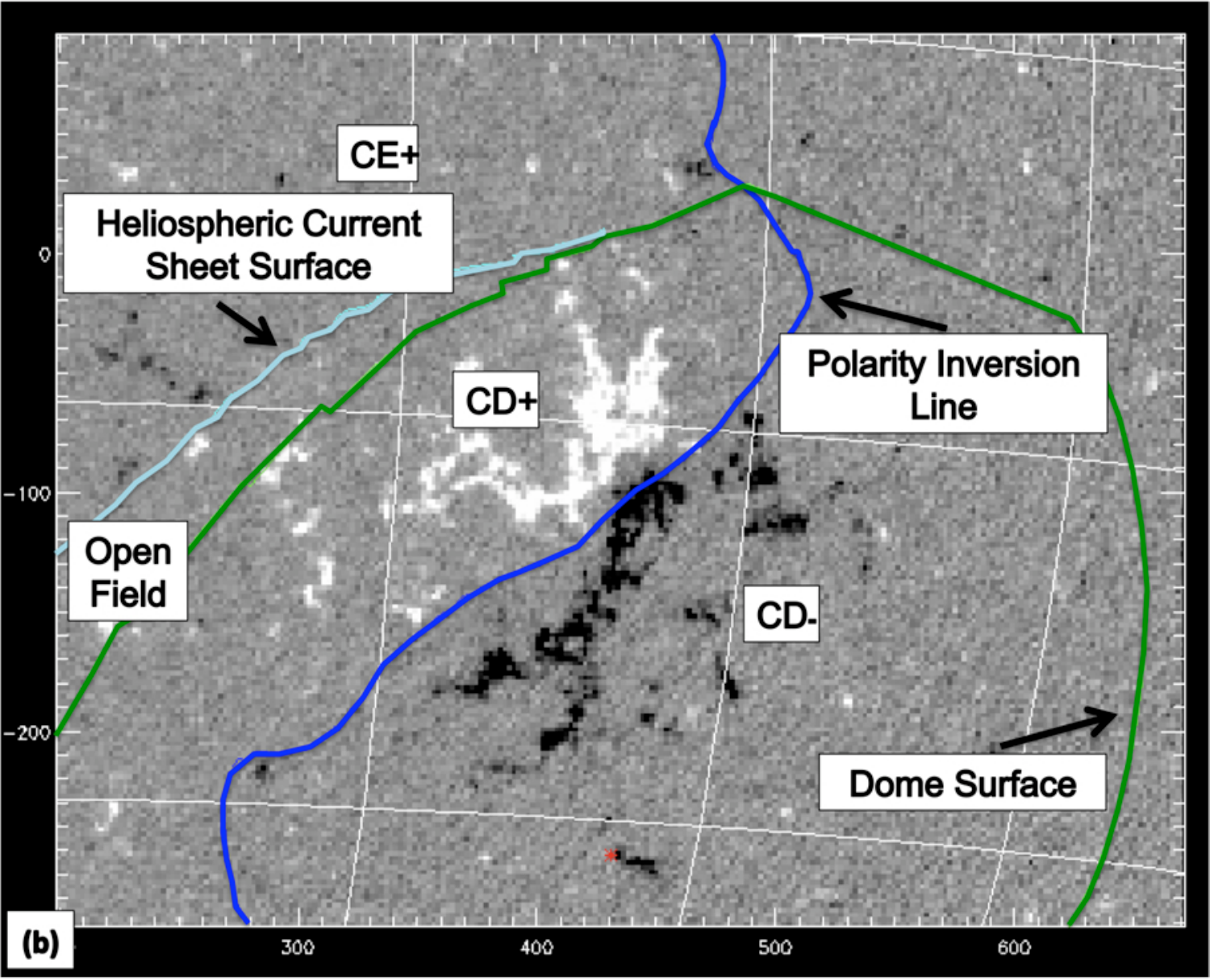}
\caption{(a)  PFSS model of the active region as it crossed the limb ($\sim$08:00 UT on 2008 April 7) using a source surface at 2.5 solar radii (black dashed circle).  \textit{Blue}:  Heliospheric current sheet and its separatrix (dashed).  \textit{Red}:  Spine lines extending from the null point (r $=$ 1.25 R$_{\odot}$) (where the spine line and dome (green) intersect).  \textit{Green}:  Dome fan surface originating from the null point which encloses all of the closed field originating from the active region's negative polarity.  (b)  Magnetogram taken a week before the flare.  CE+ is the region of positive field that closes across the equator to negative flux in the northern hemisphere.  CD+/- are the regions of positive/negative field enclosed by the dome described above.  The ``Open Field" region extends to the southeast of the active region.}
\label{magnetogram}
\end{center}
\end{figure}

The ``Cartwheel CME" flare follows this general interpretation; however, investigating the 3-D nature of this flare is important in order to understand the subtle variations from the 2-D model, namely the apparent southward progression of the CCS.  First, we invoke a PFSS model to get a general understanding of the active region's magnetic topology.  The macroscopic field is approximated using a PFSS model from the harmonic coefficients distributed by the Wilcox Solar Observatory.  The skeleton of this field is shown in Figure~\ref{magnetogram} (a) for the active region as it crossed the limb (2 days prior to the flare).   Open field crosses the source surface, located at 2.5 solar radii (black dashed circle), in regions of outward and inward flux separated by the heliospheric current sheet (HCS) mapping down to the solar surface in helmet-shaped separatrices (blue).  The flux below this surface is composed of closed field lines.  In addition there is a single coronal null point located about 170 Mm (r $=$ 1.25 R$_{\odot}$) above the active region.  The fan surface of this negative null point forms a dome (green lines) overlying all the closed field lines anchored in the negative polarity of AR 10989.

The photospheric image of the large-scale skeleton is indicated in Figure~\ref{magnetogram} (b) on top of an MDI magnetogram from one week before the flare, when AR 10989 was visible on the disk.  The image of the HCS separatrix (cyan) divides positive footpoints of the field lines which close across the equator (CE+) from open field lines.  The footprint of the dome (green) encompasses the closed field of the AR which crosses the polarity inversion line (PIL, blue).  The north-south orientation of the PIL almost certainly persists until the time of the flare, suggesting that the CS is not seen edge-on.

Based on the magnetic field topology of the active region, a scenario for this flare's evolution can be described as follows:  An eruption occurred in the field originally under the dome.  The field opened by the CME then temporarily joined the flux from the large-scale ``Open Field" region which extends to the southeast of the active region.  This conclusion is supported by SECCHI observations which show that the filament initially moves southeastward towards the ``Open Field" (see Figure~\ref{cme_path}).  Removal of closed flux would result in the dome shrinking downward with the left portion of the green curve in Figure~\ref{magnetogram} (b) likely moving to the right.  The excess open flux resulting from this shift would not be energetically favorable; therefore, reconnection began to occur (likely in the right to left direction) to replenish the field lines in the CD region and to counteract the dome shrinkage.  This reconnection is observed as shrinking loops.  The SECCHI images strongly suggest that the shrinking of loops begins in the west and progresses southeastward along the PIL.  The lifting of the dome to its initial configuration is observed as the rising of the post-eruption arcade.  Examples of an arcade brightening from end-to-end, progressively along the length of the PIL, are not new.  See, e.g., \cite{hana94} and the more recent analysis by \cite{trip06_b} referring to them as ``asymmetric eruptions."

\clearpage

\begin{figure}[!ht] 
\begin{center}
\includegraphics[width=1\textwidth]{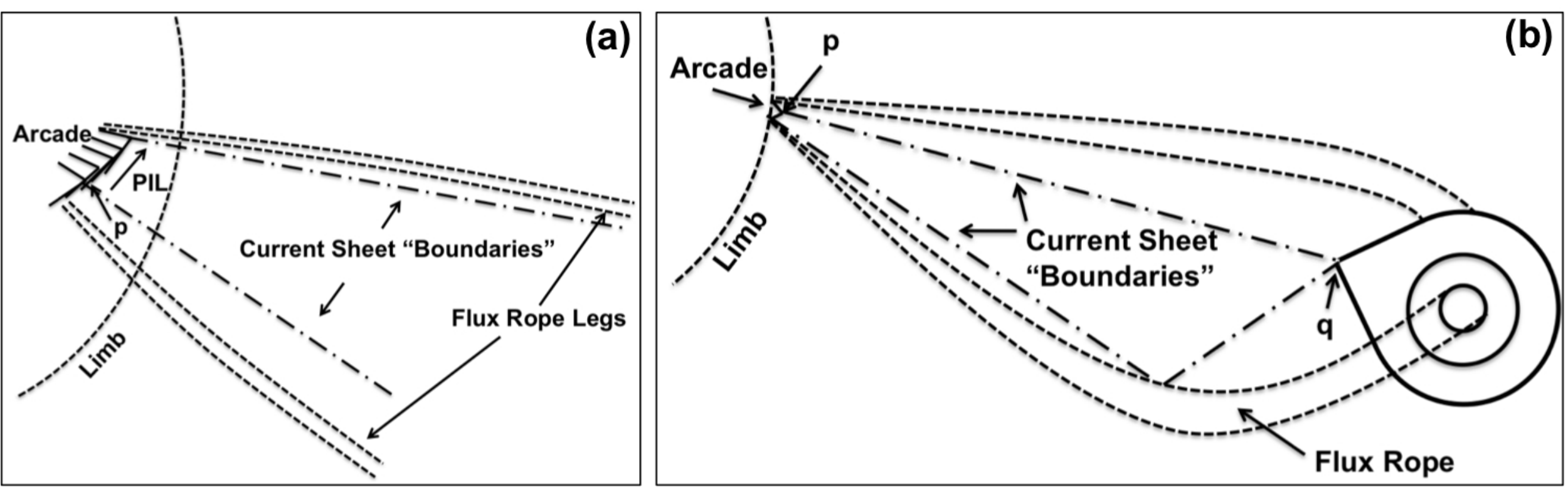}
\caption{Schematic diagram depicting a 3-D interpretation of the ``Cartwheel CME" flare as seen by (a) XRT (rotated view) and (b) LASCO.  A current sheet would form along the polarity inversion line (PIL) and between the leading edge of the erupted flux rope and arcade.  The boundaries of this current sheet could extend along the length of the arcade.}
\label{cs_cartoon}
\end{center}
\end{figure}

\begin{figure}[!ht] 
\begin{center}
\includegraphics[width=.6\textwidth]{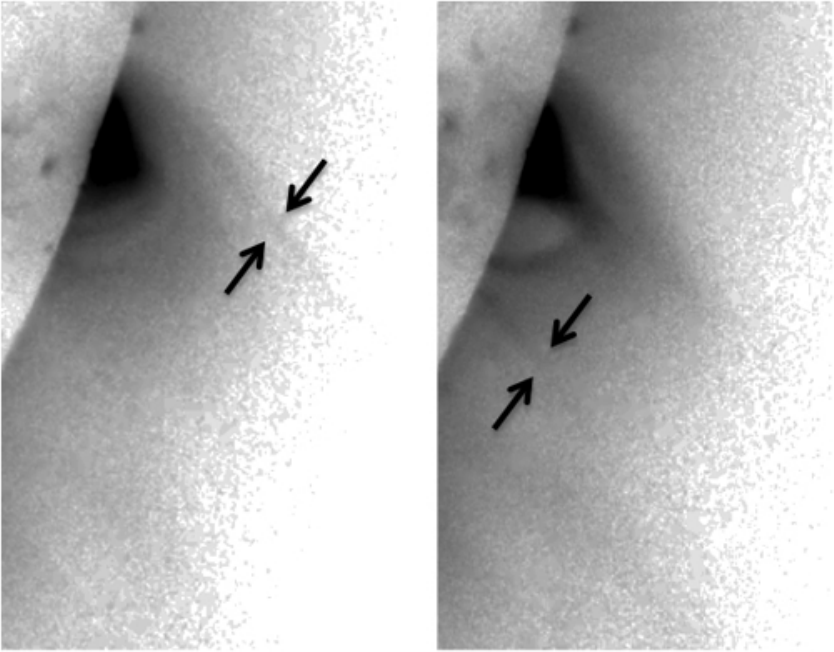}
\caption{Anchored legs of the flux rope slowly rotate southward (plane-of-sky) through the XRT FOV.}
\label{cme_legs}
\end{center}
\end{figure}

A schematic diagram for a 3-D interpretation of the ``Cartwheel CME" flare is provided in Figure~\ref{cs_cartoon}.  A FOV closely matching that of XRT, rotated so as to focus on the current sheet extending from the arcade, is shown in (a).  The outer corona (unrotated) is shown in (b) to describe the LASCO C2 images.  A current sheet would form along the PIL and between the erupted flux rope and arcade.  The current sheet ``boundaries" could extend outward from the ends of the developing arcade.  The anchored legs of the flux rope are observed by XRT to slowly rotate southward and are shown in Figure~\ref{cme_legs}.  A very similar scenario is depicted by \cite{ciaray08} (see Figure 7 therein).

\begin{figure}[!ht] 
\begin{center}
\includegraphics[width=1\textwidth]{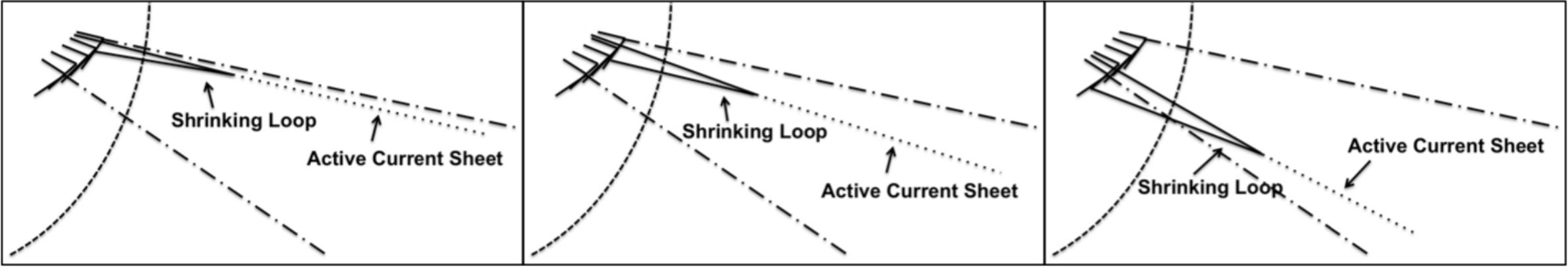}
\caption{Evolution of the current sheet near the arcade closely matching a rotated XRT FOV.  The shrinking loops begin in the west and move southeastward along the polarity inversion line.  The active current sheet indicates where reconnection is occurring and would likely appear as a bright, thin linear feature in XRT images.}
\label{cs_cartoon_sadls}
\end{center}
\end{figure}
  
\begin{figure}[!ht] 
\begin{center}
\includegraphics[width=.8\textwidth]{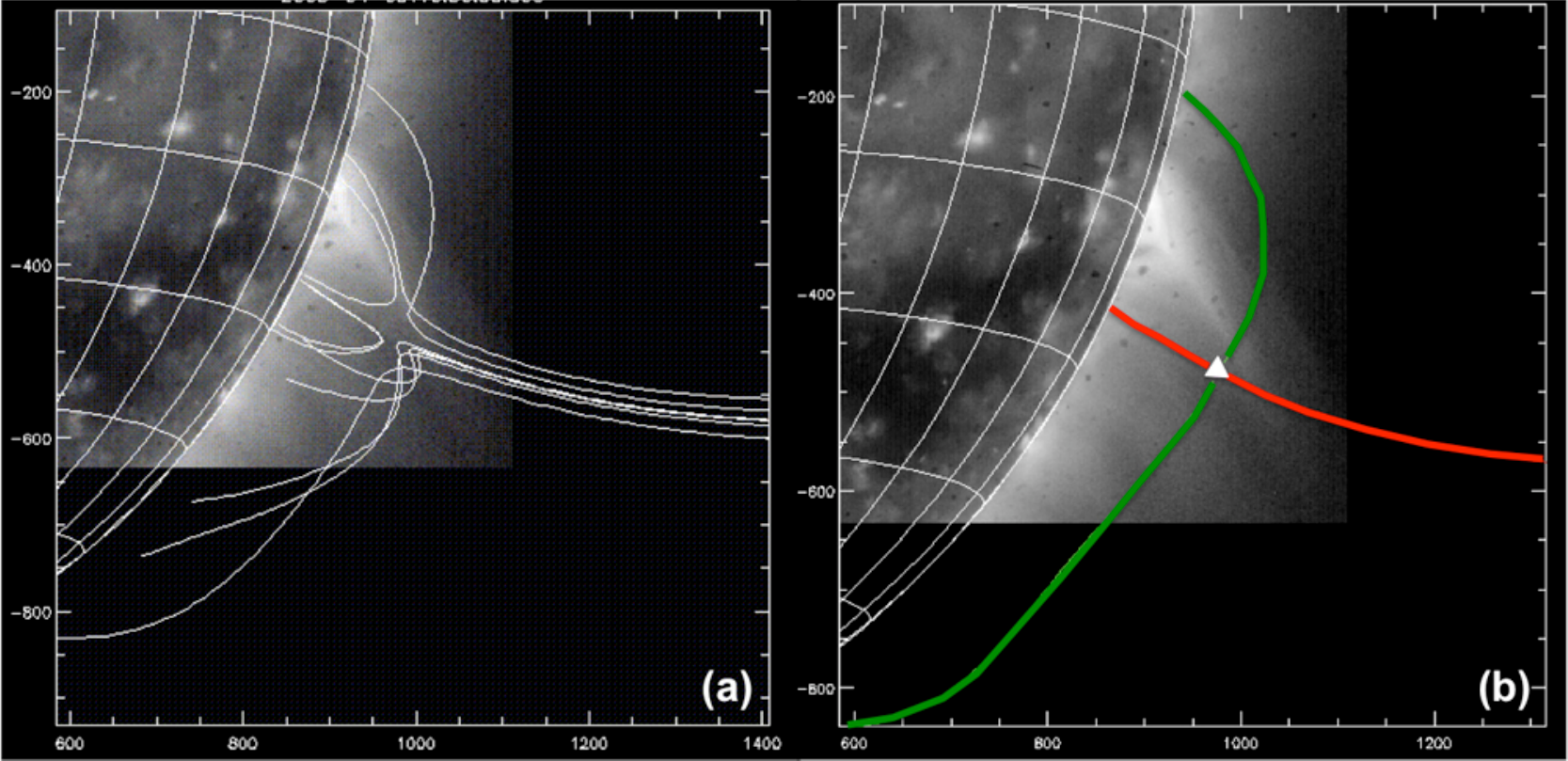}
\caption{(a)  Typical field in the neighborhood of the boundaries indicated by the colored lines in the right panel.  (b)  2-D slices of the dome and spline lines (green and red, respectively) from the PFSS model shown in Figure~\ref{magnetogram} (a).  The null point is indicated by a white triangle.}
\label{pfss_null}
\end{center}
\end{figure}
  
XRT observes high temperature plasma.  The Al/poly filter is sensitive to plasma at several millions of degrees K.  It is unlikely that the current sheet shown in Figure~\ref{cs_cartoon} would be emitting at such high temperatures except in regions of active reconnection.  We noted above that the shrinking loops appear to begin in the western region and move southeastward along the inclined PIL.  Figure~\ref{cs_cartoon_sadls} shows how the current sheet would look at different stages considering this southeast motion.  If only the active portion of the current sheet (where reconnection is occurring) were to be emitting at high temperatures \citep{reevesinpress}, then a bright, thin linear feature would be observed by XRT and appear to move southward.  This phenomenon is exactly what is observed; therefore, we propose that the current sheet is not being physically rotated.  As noted in Section 3.1, near the middle of the XRT image sequence, the CCS appears fan-like (see Figure~\ref{cs_fan}) which could indicate multiple regions of patchy reconnection.  Also, the shrinking loops early in the image sequence appear to be dark while later they appear bright.  We speculate that this may be due to an increased density available in the current sheet as the flare progresses through processes such as chromospheric evaporation.

\begin{figure}[!ht] 
\begin{center}
\includegraphics[width=.8\textwidth]{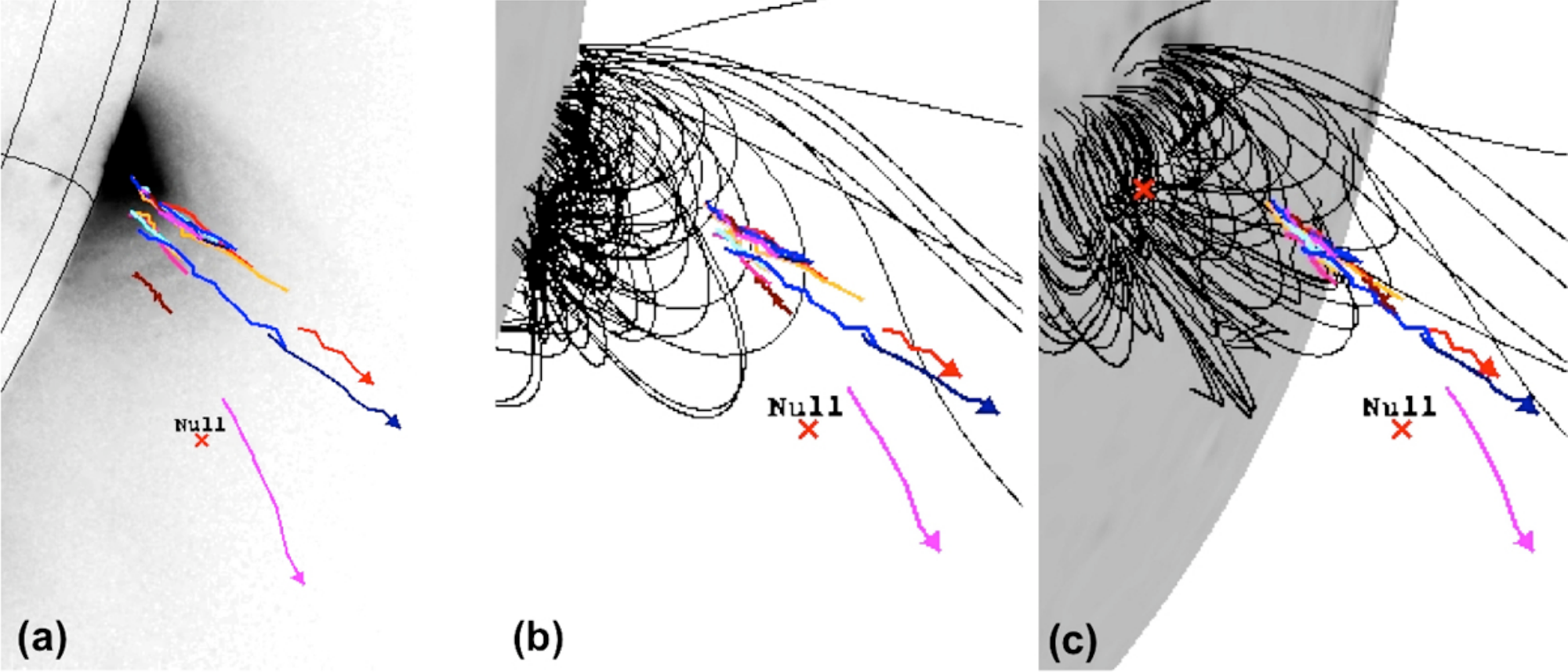}
\caption{Track composite images from Figure~\ref{tracks} with the null point overlaid.}
\label{tracks_null}
\end{center}
\end{figure}

 2-D slices of the dome and spline lines are shown in Figure~\ref{pfss_null} with the null point symbolized by the white triangle in (b).  In a 2-D model, a current sheet would be oriented along a bisector between the green and red curves in (b); however, the actual field is not potential so the location of the current sheet would not exactly match this prediction.  This null point is overlaid onto the track composite images in Figure~\ref{tracks_null}.

\clearpage

\section{Summary \& Conclusions}

Downflowing oblong voids have been observed with SXT and TRACE above arcades resulting from long-duration flaring events.  Similar flows have been observed on much larger scales with LASCO although not always in association with flare arcades (\citeauthor{mck99} \citeyear{mck99}; \citeauthor{mck00} \citeyear{mck00}; \citeauthor{mck99} \citeyear{mck99}; \citeauthor{sadsI} \citeyear{sadsI}; \citeauthor{asai04} \citeyear{asai04}; \citeauthor{khan07} \citeyear{khan07}; \citeauthor{sheeley04} \citeyear{sheeley04}).  These features, known as supra-arcade downflows (SADs), have been interpreted as the cross-sections of individual shrinking magnetic loops as they retract from the reconnection site high above the arcade.  The dark loops seen early in the image sequence are consistent with the ``shrinking empty loop" discussion pertaining to SADs in \cite{mck00} and the spectral measurements of \cite{innes03}.   The brightening of the flows in the middle of the sequence may have resulted from increased density in the current sheet as the flare progresses.

\begin{figure}[!ht] 
\begin{center}
\includegraphics[width=.7\textwidth]{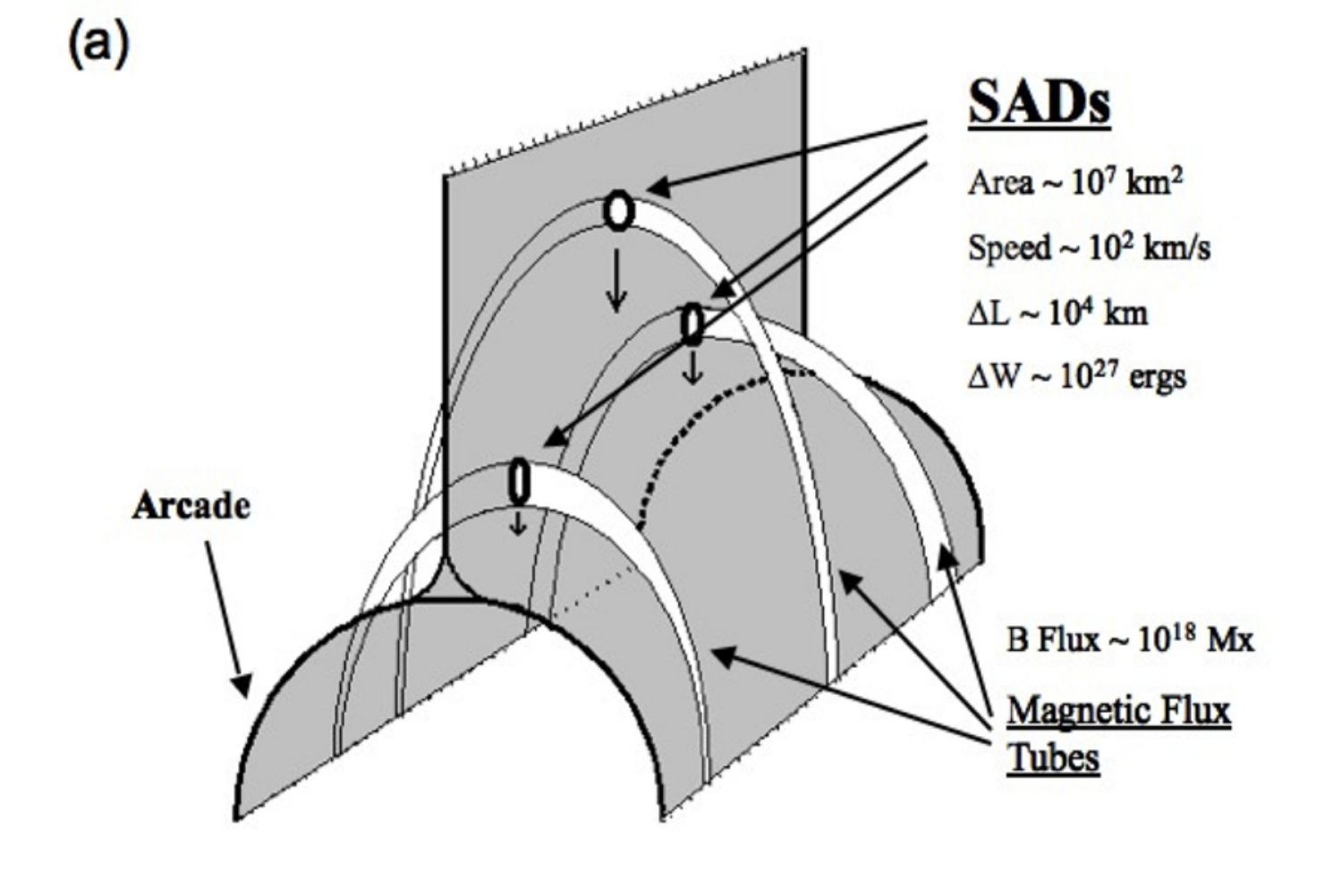}
\includegraphics[width=.7\textwidth]{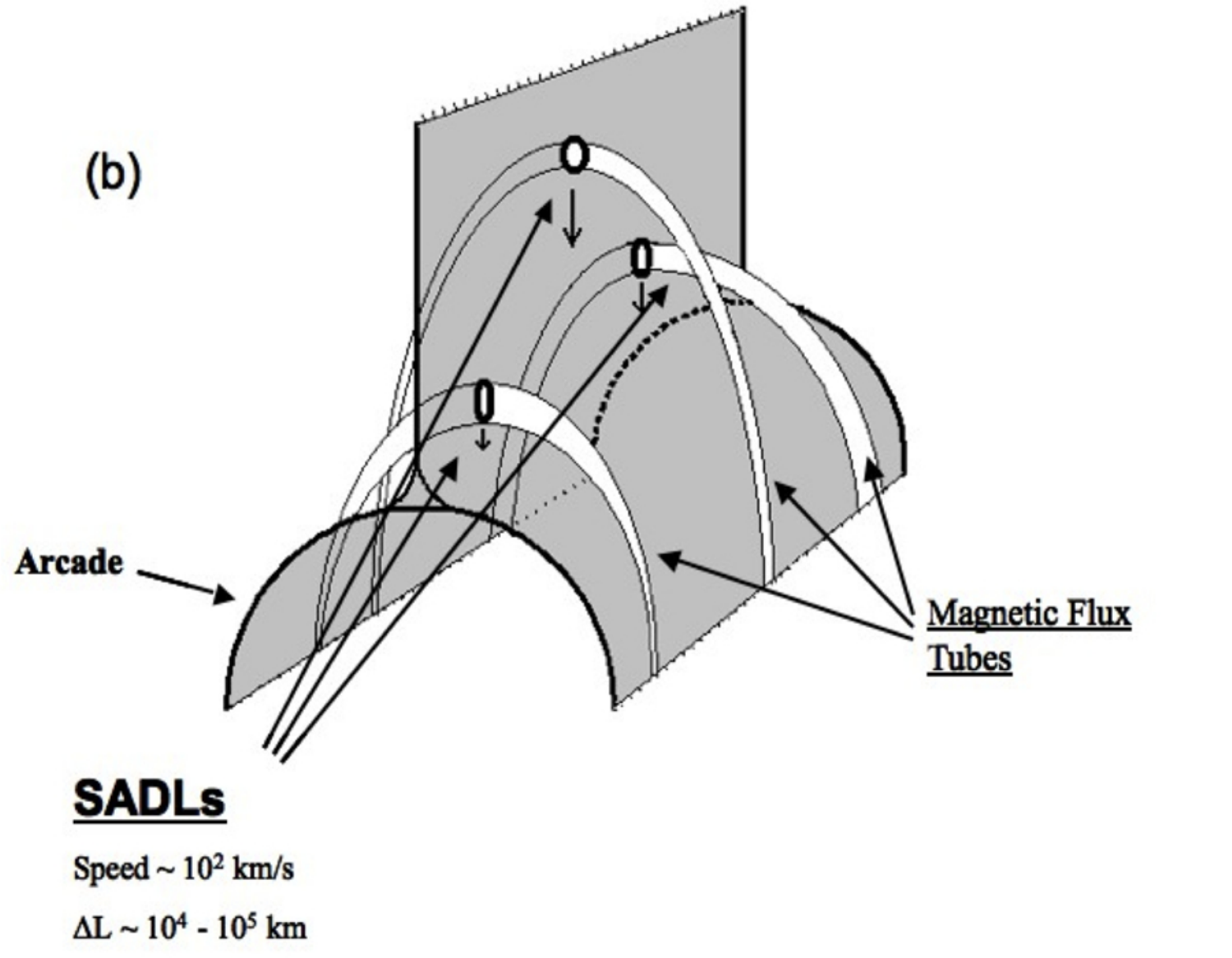}
\caption{(a)  Schematic diagram of supra-arcade downflows (SADs) resulting from 3-D patchy reconnection.  Discrete flux tubes are created, which then individually shrink, dipolarizing to form the post-eruption arcade.  The measured quantities shown are averages from the following events:  1999 January 20 (M5.2);  2000 July 12 (M1.5);  2002 April 21 (X1.5) \citep{sadsI}.  (b)  Schematic diagram of supra-arcade downflowing loops (SADLs) also resulting from 3-D patchy reconnection.  The measured quantities are averages from the flare occurring behind the western limb on 2008 April 9.  Note that the viewing angle is perpendicular to that of the SADs observations.}
\label{diagram}
\end{center}
\end{figure}

Figure~\ref{diagram}, adapted from \cite{sadsI}, gives quantitative estimates of pertinent parameters that describe this scenario based on a small sample size.  The viewing angle for SADs with respect to Figure~\ref{diagram} (a) is from the side (i.e. perpendicular to the arcade).  If this scenario is correct, then for the 2008 April 9 event, the viewing angle is face-on to the loops (i.e. along the axis of the arcade) so as to observe supra-arcade downflowing loops (SADLs) as shown in Figure~\ref{diagram} (b).  The perpendicular viewing angle is supported by the observation of shrinking loops in XRT and SECCHI as well as the magnetic field configuration derived with the PFSS modeling (see Figure~\ref{tracks} (b)).  The available relevant information obtained for this event has been labeled in the figure.

This flare is fairly unusual in that it occurred behind the limb in XRT allowing for better image quality high above the arcade with support from STEREO A observations of the flare on the limb.  Under these circumstances, a relatively faint, thin candidate current sheet (CCS) is observed in XRT and the height at which the flows are first detected is increased from that of the previously-analyzed SADs thereby increasing the overall length of the downflow path as well.  The striking similarity between the speeds and path lengths offers support to the argument that these two observational features are different views of the same phenomenon.  

Additionally, we observe a disconnection event which we interpret as a reconnection outflow pair.  We believe that this is the first clearly-observed reconnection outflow pair observed so near to the solar surface along a directly observable SXR current sheet.  The relatively narrow initial height range ($\sim$100 Mm) with respect to the CCS extent (see Figure~\ref{cs_qp_plot}) combined with the reconnection outflow pair observations seems to imply patchy reconnection with a relatively localized acceleration region.  The possible correspondence of upflows observed to propagate from within the XRT FOV into the LASCO FOV, notably the upflow associated with the LASCO CME ``pinch-off point", also support this statement.  

Reports of impulsive-phase RHESSI double coronal sources have been made by other authors (e.g. \citeauthor{liu08} \citeyear{liu08}; \citeauthor{sui04} \citeyear{sui04}; \citeauthor{sui03} \citeyear{sui03}) with the lower source corresponding to the top of the rising arcade and the upper source possibly corresponding to an ascending  reconnection outflow or an ejected plasmoid.  We note that these upper coronal sources have paths, speeds, and placements relative to the arcade similar to the bright plasmoid structure tracked as the eruption front for the ``Cartwheel CME" flare (Figure 2).  This is consistent with the extrapolated position of one coronal source from the 2002 April 15 flare \citep{sui04} which was calculated to roughly track with a coronal loop observed by LASCO (similar to Figure 16 (top)).   \citeauthor{sui04} conjecture that ``the outward-moving coronal source is part of an ejected plasmoid (or a large-scale, helically twisted loop) with two ends anchored on the Sun...".  This inference of ``an ejected ... large-scale, helically twisted loop" matches our interpretation of the 2008 April 9 eruption invoking an erupted flux rope (Figure~\ref{cs_cartoon}). 

\cite{liu08} also report that their source closer to the solar surface has a larger emission measure than the higher one.  These results are consistent with our ``Disconnection Event" observations (see Section 3.2) where the upflow portion is much dimmer and more diffuse than its downflowing counterpart.  Indeed, all of the upflows for this flare are dim compared to the bright downflows.  It is also worth noting that a non-radial, southward evolution of the loop-top source is reported for the flares in \cite{sui04} (Figure 10 therein).  The source of this divergence may have a similar mechanism as that proposed for the apparent southward drift of the CCS for this flare (Figure 20).


We interpret the basic standard picture of this eruptive flare as being initiated by the release of a flux rope by some means.  As the flux rope escapes into the outer corona, a current sheet forms in its wake.  Stretched magnetic field lines are swept together into the current sheet where reconnection occurs.  This reconnection results in the formation of pairs of cusped, looped field lines, each moving in opposite directions along the current sheet.  The loops retracting toward the solar surface form the post-eruption arcade while their companion loops are swept into the outer corona along with the erupted flux rope.  The post-eruption arcade follows the direction of the active region's polarity inversion line (PIL).  The current sheet also follows this direction as it spans the top of the developing arcade.

The ``Cartwheel CME" flare offers a unique glimpse into nearly every facet of this flaring process due to long image exposures made possible by the limb occultation of the bright footpoints.  The event is observed by several instruments (TRACE, \textit{STEREO A}/SECCHI, \textit{Hinode}/XRT, and LASCO).  SECCHI observes the onset of the flux rope eruption near the solar surface.  The formation of a candidate current sheet is seen by XRT, and we provide a possible explanation for its apparent southward progression being due to its position along an inclined PIL.  (See \cite{landi10} for additional analysis of the CME.)  Shrinking loops are very clearly seen in the XRT and SECCHI observations.  Although typically very difficult to detect due to the low signal to noise above the flaring region, XRT is able to observe flows moving outward along the current sheet with one even clearly associated with a downflow.  These upflows track into the outer corona where they appear to correspond with outflows seen by LASCO.  Finally, a post-eruption arcade develops within both the XRT and SECCHI FOV.  Any one of these observations provides an argument in favor of 3-D patchy reconnection flare models; however, all of these taken together makes the ``Cartwheel CME" flare a very compelling candidate as direct proof of reconnection.

\clearpage

\section{Acknowledgements}

This work was supported by NASA grants NNM07AA02C0 \& NNX08AG44G, NSF SHINE grant ATM-0752257, and NSF grant ATM-0837841.  The authors wish to thank Drs. L. Sui, G. Holman, J. Raymond, and W. Liu for valuable discussions and suggestions.  Hinode is a Japanese mission developed and launched by ISAS/JAXA, with NAOJ as domestic partner and NASA and STFC (UK) as international partners.  It is operated by these agencies in co-operation with ESA and the NSC (Norway).  Wilcox Solar Observatory data used in this study was obtained via the web site http://wso.stanford.edu courtesy of J.T. Hoeksema.

\end{document}